\documentclass[modern]{aastex631}
\usepackage{amsmath}
\usepackage{amssymb}
\usepackage{tipa}
\usepackage{CJKutf8}
\usepackage{wasysym}
\usepackage{combelow}
\makeatletter
\newcommand\footnoteref[1]{\protected@xdef\@thefnmark{\ref{#1}}\@footnotemark}
\makeatother

\begin{document}

\title{From Colors to Spectra and Back Again: First Near-IR Spectroscopic Survey of Neptunian Trojans}

\author[0000-0002-2486-1118]{Larissa Markwardt}
\affiliation{University of Auckland \\
SCIENCE CENTRE 303, 38 PRINCES ST \\
AUCKLAND CENTRAL, AUCKLAND, 1010, New Zealand}

\author[0000-0001-7737-6784]{Hsing~Wen~Lin (\begin{CJK*}{UTF8}{bkai}
林省文\end{CJK*})}
\affiliation{Department of Physics, University of Michigan \\
450 Church Street \\
Ann Arbor, MI 48109-1107, USA}
\affiliation{Michigan Institute for Data Science, University of Michigan \\
500 Church Street \\
Ann Arbor, MI 48109, USA}

\author[0000-0002-6117-0164]{Bryan J. Holler}
\affiliation{Space Telescope Science Institute \\
3700 San Martin Drive \\
Baltimore, MD 21218, USA}

\author[0000-0001-6942-2736]{David W. Gerdes}
\affiliation{Department of Physics, University of Michigan \\
450 Church Street \\
Ann Arbor, MI 48109-1107, USA}
\affiliation{Department of Astronomy, University of Michigan \\
1085 South University Avenue \\
Ann Arbor, MI 48109-1107, USA}

\author[0000-0002-8167-1767]{Fred C. Adams}
\affiliation{Department of Physics, University of Michigan \\
450 Church Street \\
Ann Arbor, MI 48109-1107, USA}
\affiliation{Department of Astronomy, University of Michigan \\
1085 South University Avenue \\
Ann Arbor, MI 48109-1107, USA}

\author[0000-0002-1226-3305]{Renu Malhotra}
\affiliation{Department of Planetary Sciences \\
1629 E University Boulevard \\
Tucson, AZ 85721, USA}

\author[0000-0003-4827-5049]{Kevin J. Napier}
\affiliation{Michigan Institute for Data Science, University of Michigan \\
500 Church Street \\
Ann Arbor, MI 48109, USA}
\affiliation{Department of Physics, University of Michigan \\
450 Church Street \\
Ann Arbor, MI 48109-1107, USA}

\correspondingauthor{Larissa Markwardt}
\email{larissa.markwardt@auckland.ac.nz}

\begin{abstract}

In this work, we present 0.7–5.0 $\mu$m spectra of eight Neptunian Trojans (NTs) as observed by the JWST's NIRSpec instrument. The reddest NT, 2013 VX$_{30}$, exhibits a unique spectrum with strong absorption features between 3 - 4 $\mu$m, while the bluest NT, 2006 RJ$_{103}$, shows negligible water absorption. A principal component analysis comparing these spectra with those of trans-Neptunian objects (TNOs) and Centaurs reveals that most NTs belong to the ``Bowl-type'' spectral group, while 2013 VX$_{30}$ is categorized as ``Cliff-type'' in the \citet{disco2024} taxonomy. For the bluest NT in our sample, 2006 RJ$_{103}$ shows some evidence that it may be related to carbonaceous asteroids. For the red object 2011 SO$_{277}$, we find no close TNO spectral counterpart.
Except for the true outlier 2011 SO$_{277}$, NTs have better spectral analogs among Plutinos and distant Centaurs, suggesting that spectral variation within major groups may arise from current temperature and location, rather than solely from formation regions.
Finally, we highlight optical slope (S') and near-IR slope (SIR$_1$) as effective indicators for distinguishing spectral groups and identifying outliers. These findings enable the use of broad-band photometry to explore NT and TNO surface compositions, especially for faint objects, which will be directly applicable to large photometric surveys like the Dark Energy Survey and the Vera C. Rubin Observatory's LSST.

\end{abstract}



\section{Introduction} \label{sec:intro}

Trojan asteroids are a group of small bodies that share the orbits of planets in our Solar System, librating about their two stable Lagrange points, L4 and L5. Many of them can remain dynamically stable for billions of years \citep{Lin2022, Holt2020, Gomes2016, Cuk2012, Lykawka2011}. As a result, these populations could act as a record of the early Solar System's dynamical and chemical history. Jupiter and Neptune host the largest populations of Trojan asteroids in the solar system. Both Jovian and Neptunian Trojans (NTs) have been observed to exist in thick clouds (i.e., have wide inclination distributions; \citealt{Lin2021, Lin2016, Parker2015, Sheppard2006}), and it has been suggested that these objects did not form in situ, but are instead planetesimals which were captured during the epoch of planetary migration from the same source in the primordial disk as Kuiper Belt objects \citep{Nesvorny2013, Nesvorny2009, Morbidelli2005, Kortenkamp2004}. Therefore, it follows that they should have a connection to other present day outer Solar System populations, such as Trans-Neptunian Objects (TNOs), Centaurs, and Plutinos.

The photometric distributions of such populations have been well studied, revealing that these objects have diverse, but often specifically bimodal, visible/near-IR colors. The identified TNO taxonomic groups are often referred to as ``red" or R (optical slope $\lesssim 20~\%/1000$\r{A}) and ``very red" or VR (optical slope $\gtrsim 20~\%/1000$\r{A}) \citep{Wong2017, Hainaut2012, Hainaut2002}, while Jupiter Trojans (JTs) are ``red" or R and ``less red" or LR \citep{Wong2015}. While the technical definitions (and names) of these groups has evolved, such as faint and bright near-IR in \citet{Fraser2023} and \citet{Bernardinelli2025}, the bimodality of colors has been repeatedly validated in the literature (see previous references) and is a robust result. Similarly, the Centaur population, objects with planet crossing orbits between Jupiter and Neptune, also has a bimodal color distribution similar to that of the TNOs \citep{Peixinho2003}. It was thought that the NT population was lacking VR members, which would make them unexpectedly different from the Centaur and TNO populations they fall between \citep{2018AJ....155...56J}. However, several VR NTs (2011~SO$_{277}$, 2013~TZ$_{187}$, 2013~VX$_{30}$, 2014~RO$_{74}$, 2015~VV$_{165}$) have since been identified \citep{Bolin2023, Markwardt2023, Lin2019}, and the current R to VR ratio is now similar to the TNO population \citep{Markwardt2023}. Interestingly, \cite{Markwardt2023} also identified three NTs which appear to be nearly solar in color, which may make them more similar to LR JTs. This result could be an indication that NTs have primordial or evolutionary ties to small bodies throughout the entirety of the outer Solar System, not just the TNOs.

Understanding the color distributions of these populations and how they compare to each other is important as color is often the first and most efficient measurement of bulk surface properties and composition. For example, we expect bodies in the outer Solar System to be icy and thus neutral/blue. Over time, these ices will be irradiated by the Sun, causing them to redden. However, collisions can bring pristine (i.e., blue) material back to the surface. Therefore, these color distributions can tell us the history of space weathering on small bodies in the Solar System \citep{Zhang2023, Kavuchova2012, Brunetto2006}. 
Unfortunately, this simplistic model does not consider all the independent processes that can transform the surfaces of icy/rocky bodies, including radiation from the galactic cosmic rays and charged particles in magnetospheres, ice lines, heating resurfacing events, and geologic activity, or the interplay between them \citep{Bennet2013}.

Moreover, none of these scenarios take into account the fact that primordial objects may have started out with different compositions. In fact, the existence of distinctly bimodal color distributions, rather than a continuum of surface colors, in the outer Solar System has often been cited as an indication that our primordial disk consistent of planetesimals with distinct compositions \citep{Brown2011}. Ice lines in the primordial disk are one potential explanation for stringent spectral types \citep{Wong2016}. These different primordial populations would have then been scattered throughout the solar system during planetary migration. \citep{Nesvorny2009, Gomes2005, Morbidelli2005, Nice}. This would explain the diversity we see in and across populations, but it makes tying current day colors to specific primordial populations and locations even more difficult.

One way we can try to disentangle the effects of surface ice modification, primordial composition, and dynamical evolution is to measure the surface composition of these bodies directly through reflectance spectroscopy, especially comparing such results from various populations to each other. Near-IR bands (0.5-5 $\mu$m) are especially useful for observing planetary surfaces as many broad molecular absorption features occur at these wavelengths. The NIRSpec instrument on JWST is the premier tool for making these measurements, especially since it allows for coverage of this wavelength region without telluric contamination. There are also many objects in the outer Solar System which are too faint to be observed (in a reasonable exposure time) with similar ground-based observatories. JWST observations have already been taken of various outer Solar System small bodies \citep{DePra2024, disco2024, discocentaur2024,Wong2024}. We will discuss many of these results in more detail in Section \ref{sec:comparsion}, but in general, water and CO$_2$ ice and organics have been found to be ubiquitous across these populations. This is in line with previous explanations for photometric variations being due to irradiation of such ices. Differences in these specific features have been used to create a spectral taxonomy for the largest JWST survey of Solar System small bodies \citep{disco2024}. These results also indicate that these spectral types are indeed related to their dynamical properties and evolutionary history since cold classical TNOs exclusively had organics dominated surfaces while other TNO populations exhibited a variety of compositions.

In this work, we take similar measurements using the JWST NIRSpec IFU  of eight NTs which cover a range of optical colors, including the first NT found to be VR. With this data, we hope to address the following fundamental questions: 1. Do NTs come from the same source populations as TNOs as has been presumed, and do they have links to any other populations in the outer Solar System? 2. Are color variations in the outer Solar System due to primordial compositional differences or space weathering? The NT population is key to understanding such overarching questions about the history of the outer Solar System. For one, their orbits are at an intermediate location between other distant populations. Especially regarding scattering of planetesimals during planetary migration, these objects represent a stable population between two other end states and are likely to be some mixture of what was scattered to the JT and TNO regions. Additionally, disentangling the effects of space weathering is much easier for the NT population. For one, due to their distance, these objects experience less irradiation and thermal processing than JTs and low-perihelion Centaurs. Additionally, all of the objects in our sample are stable for Gyr \citep{Lin2022, Lin2021}, meaning they have been at the same orbit, irradiation level, and temperature as each other for longer than typical resurfacing timescales. Moreover, based on absolute magnitudes (no albedo measurements have been taken for any NTs), we suspect that NTs are roughly the same size as each other and relatively small \citep[H$_r$ between 6.9 to 8.1\footnote{Diameter size is around 100 to 200 km assuming 0.1 albedo.},][]{Lin2021}, and thus geologically inactive. In other words, any differences we observe between their compositions must be due to either differences in collisional history or, more likely, primordial composition. This is \textit{not} the case for many other population, such as Centuars (especially active ones) or non-resonant and/or large TNOs. Moreover, due to the size of the stable Langrage region at the orbit of Neptune, NTs are expected to have undergone relatively fewer collisions with each other, especially as compared to the JTs \citep{bottke2023}. Therefore, their surfaces are likely to be less collisionally modified than other similar populations and could be more indicative of what the surfaces of primordial objects in this region were like. In short, this population has the potential to uniquely constrain the chemical history of the outer Solar System.

The structure of this paper is as follows: Section~\ref{sec:reduc} describes the observation and data reduction process. Section~\ref{sec:jwst} presents the spectra of our sample. In section~\ref{sec:comparsion}, we compare the NT spectra to TNO spectra and identify their closest spectral counterparts. Section~\ref{sec:colorvsspectrum} demonstrates the relationship between optical colors and the near-IR ($<1.2~\mu m$) spectra of NTs and TNOs. In this section, a schema for mapping the optical/near-IR colors to their spectral types is also presented. Finally, a brief summary of this work is given in Section~\ref{sec:summary}.

\section{JWST Data Reduction} \label{sec:reduc}

Our analysis uses the 3D data cubes created with the JWST calibration pipeline version 1.11.4 and the \texttt{jwst\_1122.pmap} reference file context. Extraction of the target spectrum from the 3D data cube can be handled via aperture photometry within each wavelength slice (i.e., within each image of the data cube). However, aperture photometry directly on the data increases the noise in the resulting spectrum; the better option is to perform aperture photometry on a scaled point-spread function (PSF) model fit to the data in order to remove artifacts and cosmic rays. 

For each wavelength slice, we constructed empirical PSF models (referred to as ``template PSFs''; \citealt{Wong2024}) using a moving median on the 10 slices shortward and longward of the considered slice (21 slices total). For slices close to the short- or long-wavelength end of the cube, less than 21 total slices were used to create the template, resulting in slightly higher noise in these regions. Use of a small wavelength window reduces variations in the background and the PSF. The background was subtracted from each slice individually prior to the calculation of the median, then the template PSF was trimmed to a 9$\times$9 pixel box (all other pixels set to NaN) and normalized to unity within the box. The template was then used as the model for a fit with two parameters: a flux scaling factor and a background. The best-fit parameters were determined by using the \texttt{scipy.optimize.minimize} Python routine to minimize the $\chi^2$ of the residual (data minus model) using a Nelder-Mead downhill simplex algorithm \citep[also known as the amoeba fitting routine;][]{2007PhRvL..98k7402P}. 

After determining the best-fit parameters, the template was multiplied by the flux scaling factor, the centroid was computed in the image, and a 3.5-pixel radius circular extraction aperture centered on the centroid was used to extract the flux. This process was performed at each wavelength to extract the full 1D spectrum from 0.7 - 5.0 $\mu$m (the noisier portions of the spectrum $<$0.7 $\mu$m and $>$5.0 $\mu$m were trimmed prior to carrying out the extraction process). The same process was also carried out for a G-type standard star (SNAP-2 from PID 1128; \citealt{2022AJ....163..267G}). The 1D spectra from all dithers of a given observation were resampled onto the wavelength grid of the standard star, normalized by the median value of each dither to account for scaling discrepancies between dithers, median-combined, and divided by the standard star to remove the solar component. Outliers were identified and removed using a 21-point moving median and a 3-$\sigma$ threshold; any points above the threshold were replaced with the value of the moving median. Uncertainties were computed as the median absolute deviation of the dithers in each wavelength bin. 

For each target, these flux ratios were then scaled by a distance correction accounting for its distinct heliocentric distance. Finally, we used a lowess (Locally Weighted Scatterplot Smoothing) function \citep{seabold2010statsmodels} to find the maximum flux ratio in smoothed spectra, which we used as a normalization value. Note, this maximum value did not occur at the same wavelength for each spectrum, but it was at about 2.5 or 3.5 $\mu$m (either side of the 3 $\mu$m feature) in all cases.

\section{Spectra of Neptunian Trojans} \label{sec:jwst}
\begin{figure}
\plotone{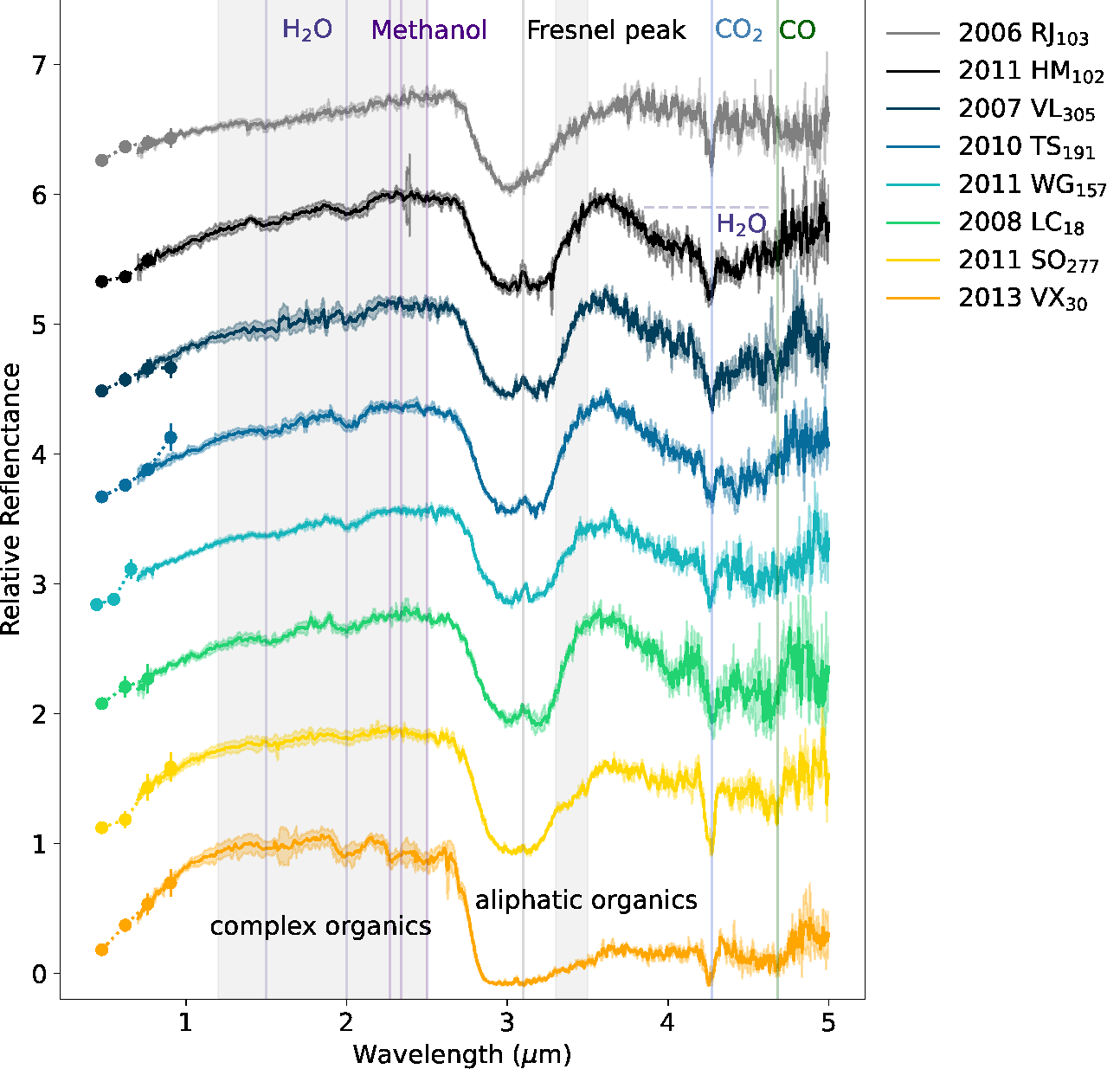}
\caption{JWST NIRSpec spectra, with arbitrary reflectance, of 8 NTs sorted by measured near-IR slope (see Sec.~\ref{sec:colorvsspectrum}) such that the bluest, 2006~RJ$_{103}$, is at the top. The spectra are shifted for clarity. Optical photometry from the literature is also plotted as solid circles (see Sec.~\ref{fig:optical+NIR}). The vertical lines, shaded areas, and dashed horizontal lines demark various molecular features evident in this wavelength region.}
\label{fig:images}
\end{figure}

The resulting spectra, sorted by our measured optical/near-IR slope (see Sec.~\ref{sec:colorvsspectrum}), are shown in Fig.~\ref{fig:images}. There are some evident features and variations in the spectra of our sample: 
\begin{enumerate}
    \item The NT which is the reddest in color (that is, it has the largest optical/near-IR slope), 2013~VX$_{30}$, also has the most distinct NIRSpec spectrum, with more absorption at wavelengths greater than $\sim$3.25 $\mu$m than any other target in this sample (see the orange spectrum shown at the bottom of Figure~\ref{fig:images}).
    \item The other ultrared NT, 2011~SO$_{277}$\footnote{\label{VR}We refer to \citet{Markwardt2023} for more nuanced discussion of this target's color classification.}, 
    has less absorption between $\sim$3.3 and 4 $\mu$m (around the region of aliphatic organics absorption) compared to 2013~VX$_{30}$ but more than the bluer NTs. Its spectrum is also generally featureless between $\sim$1.25 - 2.5 $\mu$m and has a weak Fesnel peak, similar to 2006~RJ$_{103}$.
    \item The other NTs (except for 2006~RJ$_{103}$) show an apparent 3.1 $\mu$m Fresnel peak, which is indicative of specifically crystalline water ice, and water absorption features, including broad absorption from $\sim$3.5 - 4.5 $\mu$m in all cases and 1.5 and 2 $\mu$m bands in some cases. We note that the 1.5 and 2 $\mu$m bands are least evident in the spectrum of 2010~TS$_{191}$, which has the highest amount of noise out of any of these objects in the $\sim$ 1 - 2.5 $\mu$m region.
    \item The bluest NT,  2006~RJ$_{103}$, seems to have relatively insignificant water absorption features compared to the other blue NTs. In particular, its spectrum from 1 - 2.5 $\mu$m appears to be featureless (as we observed for 2011~SO$_{277}$).
    \item All spectra show CO$_2$ absorption at 4.26 $\mu$m. This result is consistent with the recent finding of widespread CO$_2$ ices in the trans-Neptunian population \citep{DePra2024}. However, the CO absorption at 4.68 $\mu$m is not clear for most objects.
\end{enumerate}

Taking all of these observations together, we first note that despite having nearly the same present-day orbits, there is clear spectral variation in the NT population. As discussed in Section~\ref{sec:intro}, this variation is most likely due to either differences in their collisional history or primordial composition. We should note that based on previous photometric observation which suggested that NTs had different surfaces due to their color variation, spectral variation across the sample was an expected outcome. As predicted, the most extreme NT by color, 2013~VX$_{30}$, indeed has the most unique spectrum. Since highly processed organic material is a known reddening agent \citep{Zhang2023, Poston2018, DalleOre2015, Brunetto2006}, it is also not surprising that this object has the strongest organic absorption features. Meanwhile, in the overall sample, CO$_{2}$ and water ice seem to be abundant while CO is rare to non-existent. Currently, in-depth studies of the collision rates (especially compared to each other) for these objects do not exist, though \citealt{bottke2023} did suggest that NTs are likely to be less collisionally processed compared to similar outer Solar System population. Assuming collision rates are similar across the NT population, the spectral variation we observe could be an indication that these objects originated from different regions of the primordial disk but were then all scattered into the NT region during planetary migration. However, further theoretical work exploring both of these possibilities are needed to explain why NTs have their distinct surfaces.

\section{Spectral Comparison between Neptunian Trojans and other TNOs} \label{sec:comparsion}

As discussed in Section~\ref{sec:intro}, NTs and TNOs are thought to come from similar source populations in the primordial Solar System, so we would expect them to have similar bulk compositions. However, there is an obvious inconsistency between our results and similar TNO observations in that we find no robust detection of CO in the NT population even though it was found to be common in TNOs \citep{DePra2024}. However, the vapor pressure of CO is high ($\sim 10^{-4}$ Pa at 29K) and sensitive to temperature in the trans-Neptunian region \citep{Grundy2024, Millan2024}. Since the black-body temperature at 30 AU is about 50K, which is 10K higher than it is at 40 AU, if the NTs once had CO ice on their surfaces (as observed on the TNOs), it may have vaporized at this point considering the NTs have resided in their current orbits for billion years. CO$_2$ on the other hand has a much lower vapor pressure \cite[$\sim 10^{-4}$ Pa at 87K,][]{Millan2024}, so the presence of CO$_2$ on every NT, similar to TNOs, is still expected. However, \citealt{Henault2025} suggested that the CO observed on TNO surfaces is very likely to be a secondary product of either CO$_2$ or methanol irradiation and therefore not related to the composition of primordial source populations. Since we have observed CO$_2$ on the NT surfaces, one may then expect to also observe new CO as a result of irradiation of that CO$_2$. However, NTs are still at a temperature of 50K now, so even if the CO$_2$ on their surfaces is currently being irradiated and producing CO, that CO should still be vaporized and thus not be present in our observations (which is completely consistent with the experimental results in \citealt{Henault2025}).
Therefore, we would \textit{not} rule out a common primordial source population for both TNOs and NTs based on these observations; instead it is likely an indication of the difference in thermal processing and physical sizes between these two populations.

\subsection{Principal Component Analysis}

\begin{figure}
\plotone{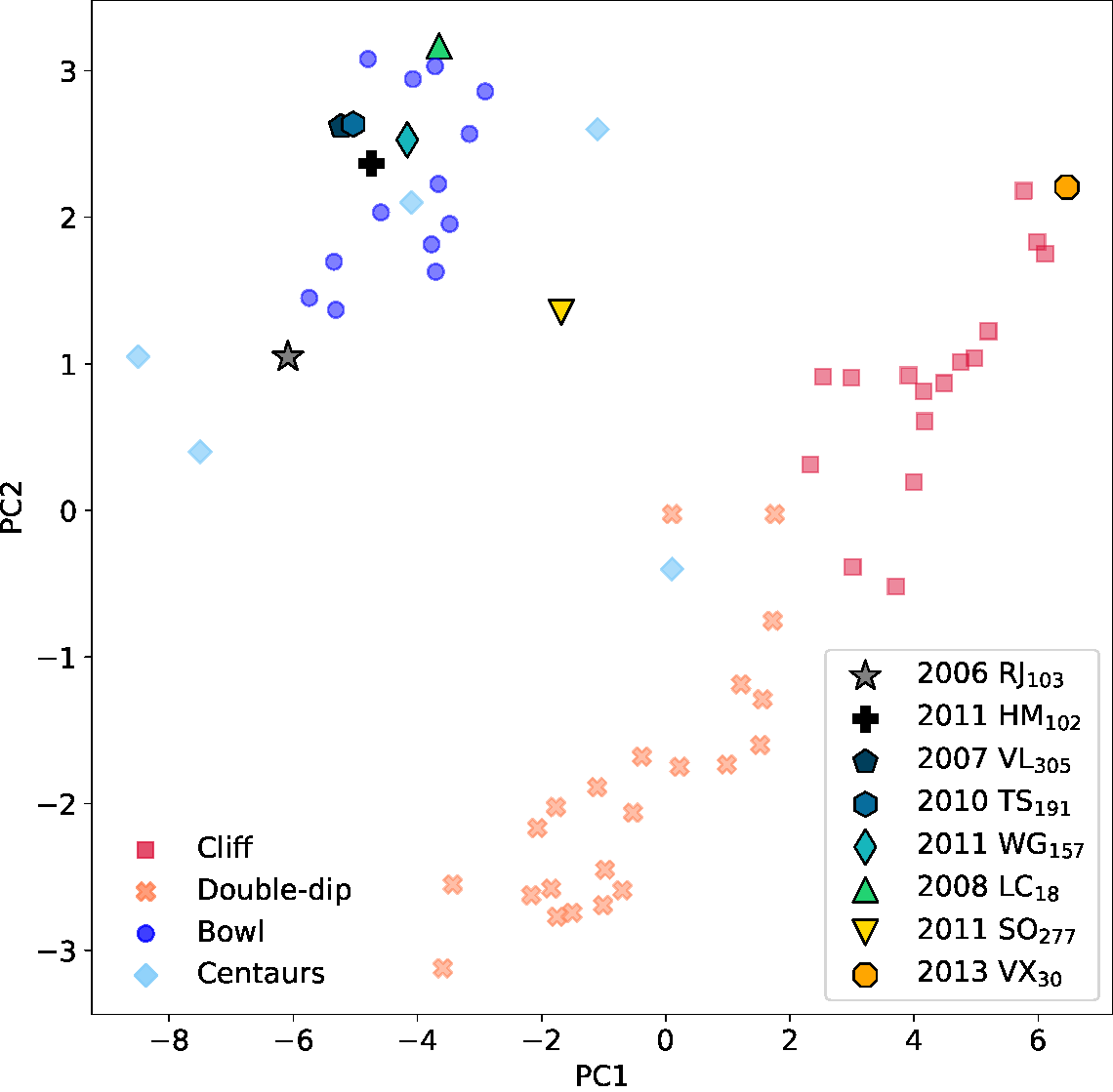}
\caption{Results of a PCA to reduce the dimensionality of the full NIRSpec spectra of NTs, Centaurs, and TNOs. The TNOs have previously been grouped into three different spectral types (Cliff, Double-dip, and Bowl) using an independent PCA \citep{disco2024}. The Centaur data are from \citet{discocentaur2024}.}
\label{fig:PCA}
\end{figure}

Further comparisons can be made between the NT and TNO populations using the full NIRSpec spectra. In particular, the TNO spectra have already been grouped into three major types based on feature inspection and clustering methods: Cliff, Dobule-dip, and Bowl-type \citep{disco2024}. To directly compare and analyze NT spectra against these categories, we implemented a Principal Component Analysis (PCA) using \texttt{sklearn.decomposition.PCA} from the \texttt{scikit-learn} package \citep{scikit-learn}. PCA is a linear dimensionality reduction technique – essentially, it breaks down a multivariate dataset into a set of orthogonal components, capturing the maximum variance. Since the TNO spectra in \citet{disco2024} have a slightly different wavelength range (0.75 - 5.1$~\mu$m), compared to the NT spectra (0.7 - 5.0$~\mu$m), we first ran the PCA with those 51 TNO spectra cut to 0.75 to 5.0$~\mu$m, and then used that result to project the NT spectra (using the same wavelength range) into the PC1 and PC2 space. The result is shown in Figure~\ref{fig:PCA}, and the values of PC1/PC2 for the NTs are listed in Table~\ref{Tab:slopes}. Although the values of PC1 and PC2 for the TNOs are slightly different from those in \citet{disco2024}, the general distribution and the relative positions of each spectrum in the projection space are still very similar to the original result.

The PCA result shows that the majority of our NT spectra belong to the Bowl type in the \citet{disco2024} taxonomy, which are dominated by water ice spectral features. On the other hand, the reddest NT, 2013~VX$_{30}$, is clearly most similar to the Cliff type TNOs as they share the same ``cliff'' shape in the 3$~\mu$m absorption band (i.e., very strong absorption from $\sim$3.5 - 4 $\mu$m). The second reddest NT, 2011~SO$_{277}$, is a clear outlier in this projection (see Fig.~\ref{fig:PCA}); it lies in between the three main groups and we would not classify it as being in any of these spectral types. Finally, the bluest NT, 2006~RJ$_{103}$, falls at the edge of the TNO Bowl group, likely due to its lack of their characteristic water absorption features. Therefore, this NT is also an outlier in the TNO taxonomy. Interestingly, we find no Double-dip types in our sample despite them being the most common (nearly half) in the original TNO sample \citep{disco2024}.

\begin{table}[ht]
	\caption{Optical/near-IR Linear Slope and Principal Components of NT Spectra}
	\centering
	\begin{tabular}{l c c c c}
		Object &  S' ($\%/1000$\r{A})  & SIR$_1$ ($\%/1000$\r{A}) & PC1 & PC2\\
		\hline
		2006 RJ$_{103}$ & ~7.87 $\pm$ 3.89 & 3.35 $\pm$ 0.11 &  -6.09 & 1.05\\
		2007 VL$_{305}$ & 10.22 $\pm$ 4.60& 5.76 $\pm$ 0.10 &  -5.23 & 2.62\\
		2008 LC$_{18}$ & ~9.96 $\pm$ 6.80 & 6.04 $\pm$ 0.14 &   -3.65 & 3.17\\
		2010 TS$_{191}$ & 11.87 $\pm$ 2.88 & 5.37 $\pm$ 0.07 & -5.04 & 2.63\\
		2011 HM$_{102}$ & ~6.47 $\pm$ 3.59 & 5.40 $\pm$ 0.09 &  -4.74 & 2.37\\
		2011 SO$_{277}$ & 17.97 $\pm$ 5.93 & 6.97 $\pm$ 0.02 &  -1.68 & 1.35\\
		2011 WG$_{157}$ & 15.37 $\pm$ 5.2 & 5.07 $\pm$ 0.07 & -4.16 & 2.53\\
		2013 VX$_{30}$ & 44.95 $\pm$ 4.79 & 9.94 $\pm$ 0.20 &  ~6.46 & 2.20\\
	\end{tabular}
	\label{Tab:slopes}
\end{table}

\subsection{Closest Counterparts}
\begin{figure}
\plotone{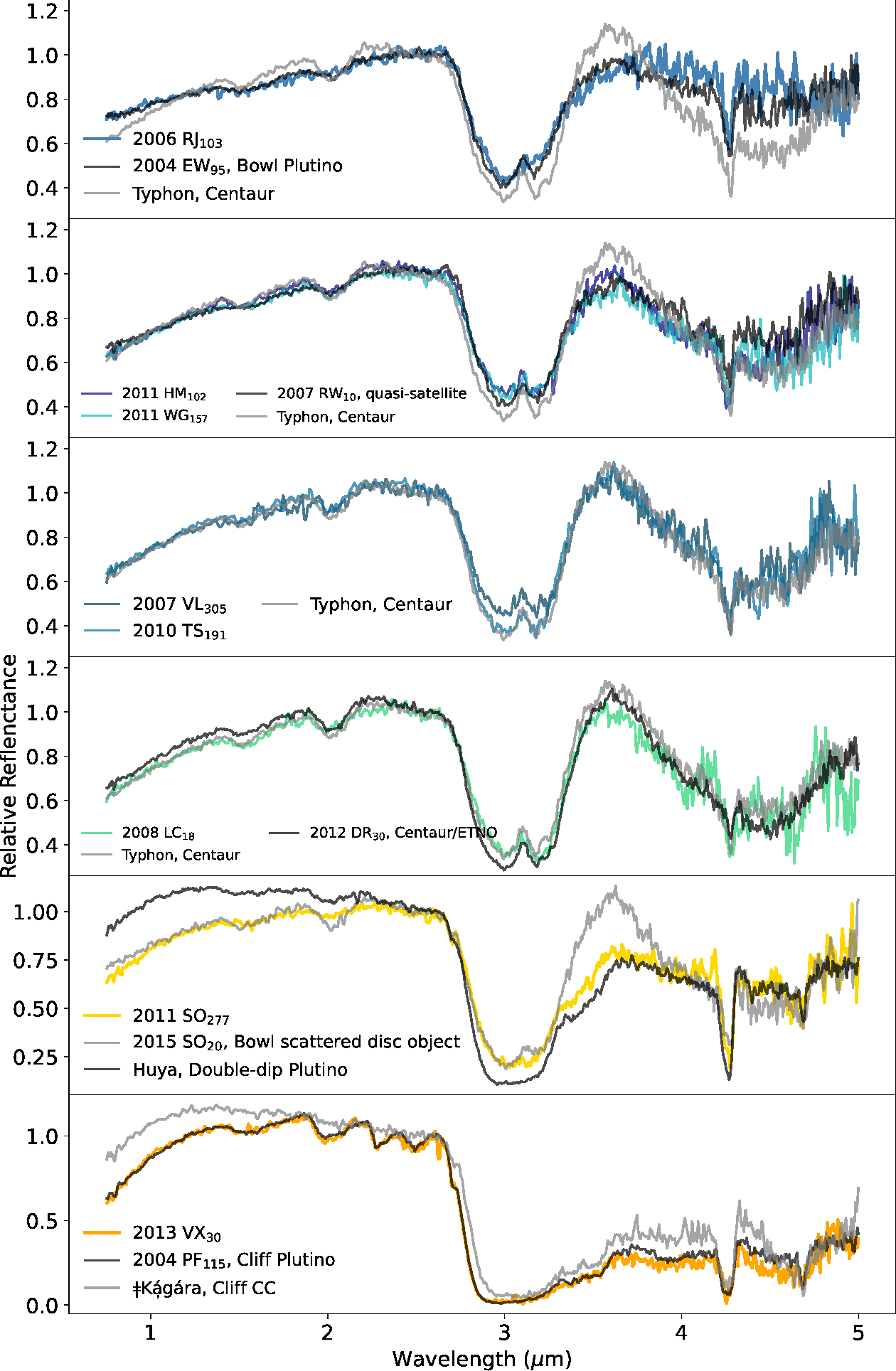}
\caption{The spectra of NTs (colored) and their counterparts (grey and black), based on minimizing the sum of the squares of the residuals between the NT and TNO spectra.}
\label{fig:comp}
\end{figure}

Since most of the NTs do seem to fit into the TNO spectral taxonomy, we then further search for specific targets in the TNO sample which are the closest counterpart to each NT spectrum. We calculate the sum of the squares of the residuals between the NT and TNO spectra to find the closest counterpart. The results are shown in Figure~\ref{fig:comp}.

\subsubsection{The Bowl-type Neptunian Trojans}
The five bowl-type NTs (2011~HM$_{102}$, 2011~WG$_{157}$, 2007~LV$_{305}$, 2010~TS$_{191}$, and 2008~LC$_{18}$) have close counterparts from the scattered population Bowl-types, such as 42355~Typhon, 2007~RW$_{10}$, and 2012~DR$_{30}$. We notice that these TNOs are not typical scattered disk objects and are sometimes referred to as \textit{Centaurs} by the definition that their perihelion is less than 30AU but greater than 5AU. However, none of them would be considered Centaurs by the stricter definitions from the Minor Planet Center\footnote{See: \url{https://www.minorplanetcenter.net/iau/Unusual.html}} and JPL SSD\footnote{See: \url{https://ssd.jpl.nasa.gov/tools/sbdb_query.html}} which limit the semi-major axis to be within Neptune's orbit. Therefore, we consider all of these counterparts to be ``distant" Centaurs, with the exception of 2007~RW$_{10}$, which is known to be a quasi-satellite of Neptune and therefore has a similar orbit to NTs \citep{2007rw10}. While these definitions may seem pedantic, they are important to the underlying question of whether surface processing is the cause of spectral variations on the surfaces of small bodies. In particular, the amount of thermal processing and irradiation these objects experience depends on how close they get to the Sun and the amount of time they spend in this more intense environment. Therefore, it should not be surprising that multiple NTs are most similar to the quasi-satellite 2007~RW$_{10}$ as they live in a very similar orbit (though it does have a perihelion nearly 10 AU smaller than these NTs).

The ``distant" Centaurs on the other hand, have perihelia much smaller than any NT, with 2013~LU$_{28}$ having the smallest perihelion distance (within the orbit of Saturn). Therefore, one may expect these objects to be further depleted in volatiles due to their closer passages to the Sun. Indeed, the crystallization of water ice (which we can see is present on these surfaces) due to higher temperatures and subsequent release of trapped volatile gas is thought to be a key driver of activity on Centaurs \citep{Jewitt2009AJ}. However, the large semi-major axes of these particular Centaurs means that they likely have had relatively few close approaches to the Sun; therefore, their surfaces may actually be less thermally processed than more typical Centaurs which could explain their apparent similarities to the Bowl NTs. Careful modeling and measurements of the water ice features, especially the Fresnel peak, 1.5, and 2 $\mu$m features, to determine the degree of crystallization is likely key to understanding the thermal processing of these objects, but is beyond the scope of this work. For now, it is clear that crystalline water ice is abundant in both the NT and Centaur populations. Since the thermal history of these targets should be somewhat different, this likely points to a common origin for these members of different scattered populations.

\subsubsection{A Cliff-type Neptunian Trojan}
It is also no surprise that the counterpart to the very red 2013 VX$_{30}$ is in the Cliff type TNOs. In Figure~\ref{fig:comp}, we compare 2013 VX$_{30}$ to two distinct counterparts, a Cold Classical TNO, (469705) 2005~EF$_{298}$ \textdoublebarpipe K\cb{\'a}g{\'a}ra and a Cliff-type Plutino, 2004 PF$_{115}$. The Cold-Classical TNOs were unique in the \citet{disco2024} sample in that they were exclusively classified as Cliff types \citep{Souza-Feliciano2024}, while all the other subpopulations were comprised of varying combinations of the three; this result supported the idea that Cold Classical TNOs have a unique dynamical history in that they have remained at their formation location.

However, there are several clear differences in the spectral features of \textdoublebarpipe K\cb{\'a}g{\'a}ra compared to 2013 VX$_{30}$; namely, \textdoublebarpipe K\cb{\'a}g{\'a}ra appears to have a less red spectral slope below $\sim$1.5 $\mu$m, is spectrally featureless between $\sim$1.5 - 2.5 $\mu$m, has a 3 $\mu$m feature which is shifted to comparatively longer wavelengths, and has less absorption between $\sim$3 - 4.5 $\mu$m. In short, while both have the distinctive 3 $\mu$m cliff shape, a detailed comparison of other features shows that they must have somewhat different surfaces. Several of these features point to 2013~VX$_{30}$ having more organics on its surface. Since Cold Classical TNOs have more distant orbits, this could be an indication that they are the same spectral type, but 2013~VX$_{30}$ is more irradiated. However, 2013~VX$_{30}$ also seems to have water or organics features which are not apparent on \textdoublebarpipe K\cb{\'a}g{\'a}ra, specifically absorptions between 2 to 2.6 $\mu$m. If it does have water ice on its surface, the shift of the left edge of its 3 $\mu$m feature could be an indication that it is predominately in an amorphous state \citep{2009ApJ...701.1347M}. While crystalline water ice features (i.e., the Fresnel peak) have been commonly observed in these populations, radiation can re-amorphize at temperatures below 100K \citep{Fama2010, Leto2003, Baratta1991, Kouchi1990}. Therefore, all of these differences seem to point to the same conclusion: 2013~VX$_{30}$ is an especially irradiated object (as we had predicted based on its VR color).

On the other hand, the spectrum of the Plutino 2004 PF$_{115}$ is a much better counterpart, being nearly identical to 2013 VX$_{30}$. Considering that Plutinos and NTs are both resonant populations which are expected to have been captured into their present-day orbits due to planetary migration, these results support the idea that the dynamically Cold Classicals are indeed a distinct population from other TNOs. We refer to \citet{Brunetto2025} for detailed studies of subpopulations of cliff-type spectra and the possible reasons of the spectral diversity.

\subsubsection{A Carbonaceous Neptunian Trojan?}
In our sample and the TNO spectral taxonomy, 2006~RJ$_{103}$ is an outlier due to its apparently dampened water absorption features. However, we are able to find a Bowl-type counterpart for it, the Plutino 2004~EW$_{95}$. While this Plutino has water features which are a bit stronger than those of 2006~RJ$_{103}$ (though one can see that both are dampened compared to Bowl-type Centaur Typhon in Figure~\ref{fig:comp}), their bluer spectral slopes are especially well matched.

In previous studies, 2004~EW$_{95}$ has been suggested to be a carbonaceous asteroid in the Kuiper Belt that still retained its primordial water ice \citep{Seccull2018}. Based on an optical spectrum of 2006~RJ$_{103}$, \citet{Sharkey2023} reported a likely 0.7 $\mu$m absorption feature due to Fe$^{2+}$ charge transfer features in hydrated silicate materials. Therefore, one interpretation of our results is that both objects were originally carbonaceous asteroids with 2006~RJ$_{103}$ now being slightly more hydrated while having slightly less water ice. If the surface of 2006~RJ$_{103}$ is indeed more hydrated, this could be due to some amount of melting of water ice on its surface either due to exposure to higher temperatures or collisions with other small bodies or micrometeorites.
It is also important to note that while irradiation does redden the surfaces of icy bodies (see more in Sec.~\ref{sec:intro}), further irradiation of hydrocarbons can actually lead to loss of hydrogen, leading to spectra which are featureless and more neutral in color \citep{1987JGR....9214933T, 1998Icar..135..389C}.
Albedo observations would also be useful here because while the colors of these objects may go from neutral to red to neutral again, their albedo should continuously darken as their surfaces are irradiated. Unfortunately, no albedo measurements of NTs have been taken to date. However, in their absence, neutral carbon-dominated surfaces due to solar irradiation are at least consistent with observations these two bodies.

All that being said, the question still remains: why is 2006~RJ$_{103}$ different from other \textit{NTs}? All of our targets are unlikely to have been recently captured, meaning they have been in very similar orbits for billions of years. Therefore, 2006~RJ$_{103}$ should not be significantly hotter or more irradiated than any other NT. One possibility is that 2006~RJ$_{103}$ has experienced more surface collisions for some reason, which could lead to hydration of its surface and potentially revealing fresh, blue ice under the irradiated crust; albedo measurements would be especially useful here as this fresh ice should also be quite bright. Another possibility of such collisions is that they instead excavated hydrated material from this object's subsurface, rather than producing it directly (as is discussed in \citealt{bottke2023}). However, as we have mentioned, the NTs are relatively small in size ($\lesssim$ 200 km), so it is not clear if thermal processing and thus aqueous alteration and differentiation of their interiors is possible.

The other possibility is, again, that 2006~RJ$_{103}$ was captured from a different source population than that of most NTs, likely from the more inner Solar System. This may also explain why this spectral type is rare in TNOs in general as closer proximity would make it easier to capture these objects into the NT and Plutino regions. Dynamical models of the transport of primordial material throughout the Solar System during planetary migration will be key to testing this hypothesis. However, assuming it is correct, the CO$_2$ observed in these spectra would need to be obtained during or after the capture process (i.e., it could not primordial), because the original source region would likely be too hot for CO$_2$ ice. Subsequent capture of CO$_2$ could also explain why there are no Double-dip NTs. Double-dip TNOs have particularly strong CO$_2$, and most are in distant TNO populations (with only two Double-dip Plutinos; see Figure 1 in \citealt{disco2024}). Therefore, in this scenario, the NT (and, to a lesser extent, Plutino) population was more likely to be populated with objects originating from closer in regions of the primordial Solar System, potentially in front of a CO$_2$ snow line, and only captured CO$_2$ on their surfaces after settling in the cooler outer Solar System. Again, dynamical models which try to recreate these compositional observations are needed to confirm this idea.


\subsubsection{The Outlier}
Finally, as expected, we do not find an obvious counterpart for the clear outlier, 2011~SO$_{277}$. We find that a Bowl-type scattered disc object, 2015~SO$_{20}$, matches the 2011~SO$_{277}$ spectrum reasonably well until $\sim$3.25 $\mu$m (though 2011~SO$_{277}$ does have a redder spectral slope at the short wavelength end). Beyond $\sim$3.25 $\mu$m, the spectrum of 2011~SO$_{277}$ fits better to that of the Plutino Huya. Indeed, the CO$_2$ absorption at 4.25 $\mu$m of 2011~SO$_{277}$ is the strongest among NTs, consistent with Huya's classification as a Double-dip type (although Huya notably has weak CO$_2$ absorption compared to other TNOs in \citealt{disco2024}). Without a clear counterpart (and more sophisticated chemical modeling which we leave for future work), 2011~SO$_{277}$ is a bit of an enigma. While they have much more similar orbits, the differences with Huya may not be surprising as Huya is a known binary \citep{huya2012} and interactions between its members could affect its composition.

2015~SO$_{20}$ on the other hand has a much more distant orbit than either of these objects, so its surface should be more pristine. Could 2011~SO$_{277}$ then be a (partially) processed Bowl-type? Indeed, even when just compared to the whole NT sample by eye, 2011~SO$_{277}$ does look transitionary, potentially between the typical Bowl-type NTs and 2006~RJ$_{103}$ due to its weak water features but stronger aliphatic organic absorption. Its stronger CO$_2$ could further support this idea. A subset of Centaurs are known to be active, exhibiting a dispersed coma of gas and dust, and the active Centaur Chiron was also observed by JWST to have strong CO$_2$ absorption bands \citep{2024A&A...692L..11P}. In fact, the prevailing theory for the cause of gas activity on these objects is the crystallization of amorphous ice, which releases volatile gases trapped in the amorphous ice pores, when these objects are at higher temperatures nearer the Sun (see \citet{2015AJ....150..201J} and references therein). We should note that Chiron specifically is not expected to reach temperatures high enough for water ice crystallization \citep{2024A&A...692L..11P}. However, low perihelia and thermal processing do seem to be related to the neutralization and paucity of VR Centaurs in some way \citep{2019AJ....157..225W}.  2011~SO$_{277}$ is 
VR\footnoteref{VR}, and if it is thermally processed and potentially active (though extended emission has not been observed to date), one would expect its water ice to be crystalline (especially as this is the likely driver of activity). However, 2011~SO$_{277}$ only has a hint of a Fresnel peak which is not clearly distinguishable from noise (again, more sophisticated modeling of this feature is left to future work). Considering its apparent lack of 1.5 and 2 $\mu$m water absorption, this object may just be depleted in water, similar to 2006~RJ$_{103}$. The difference between these two objects would then be the amount of aliphatic hydrocarbons (such as methane and ethane) and their spectral slopes near 1 $\mu$m. As we mentioned previously, the irradiation of hydrocarbons can actually lead to loss of hydrogen and a carbon-dominated surface. Therefore, while 2006~RJ$_{103}$ could represent the end state of such irradiation (neutral and spectrally featureless), 2011~SO$_{277}$ could represent the transitionary state (red and covered aliphatic hydrocarbons) from the starting Bowl-types (neutral and water-rich).

Assuming this evolution is correct, the question still remains: \textit{why is 2011~SO$_{277}$ the only transitionary object we have observed?} This is especially considering that many Centaurs pass closer to the Sun than the NT region and should be similarly processed. One explanation is that this particular spectral state is very short lived, making it simply less likely to be captured in observations. 
Assuming that it is indeed short-lived, it is possible that 2011~SO$_{277}$ was a transitionary object \textit{during} the period of planetary migration and was subsequently captured into the NT region where its spectral state has since been ``locked in" for billions of years. Unfortunately, there is nothing, including its orbit or dynamical evolution, that distinguishes 2011~SO$_{277}$ from other NTs to provide further clues. More spectral observations of distant small bodies will lead to a more complete picture of the possible types of surface compositions of this region and will allow us to determine if 2011~SO$_{277}$ is a true outlier. Observations of objects which should be minimally processed since the epoch of planetary migration may prove to be particularly elucidating. 

.
\section{Optical Colors to Spectra Relation} \label{sec:colorvsspectrum}

In previous sections of this paper, we have demonstrated the value of spectral observations (especially those taken with JWST) to determine and compare the surface composition of various small bodies in the outer Solar System. However, the TNOs and NTs in these samples are the brightest of their respective populations, and many objects are far too faint for similar observations with currently available facilities. In this section, we investigate the possibility that the spectral type of these objects (particularly the NTs) can still be inferred by using only their optical colors which can be measured with less exposure time on less competitive facilities or massively obtained by multicolor sky surveys.

\begin{figure}
\plotone{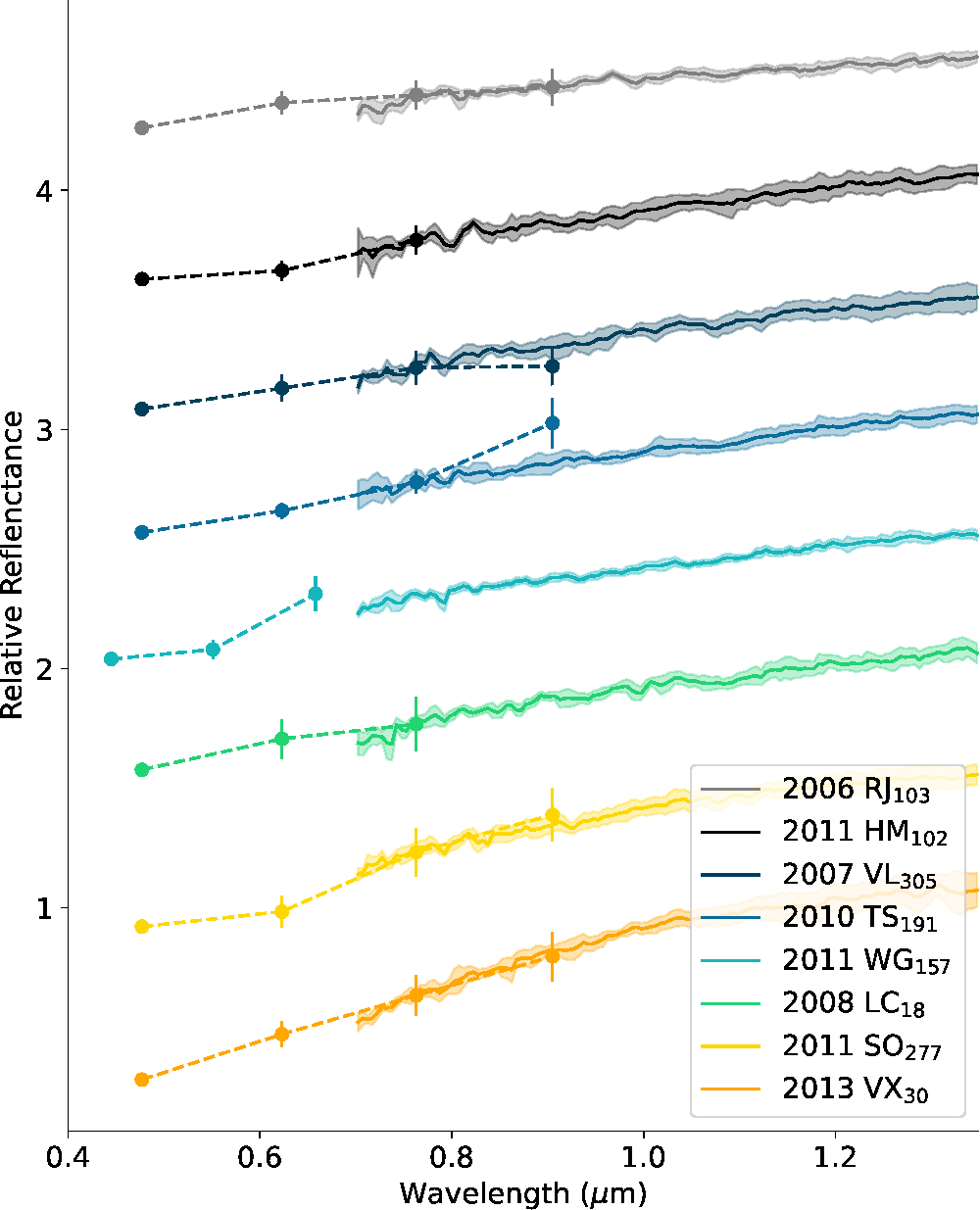}
\caption{The joint optical photometry and NIRSpec spectra of eight NTs. Each plot color corresponds to a different NT, while the circular markers correspond to the central wavelength of different optical bands (from short to long wavelength: g, r, i, z or  B, V, R for 2011~WG$_{157}$). The error-bar and the shaded regions represent the 1 sigma uncertainties on the optical photometry and NIR spectra respectively.
References for the optical photometry data: 2011~WG$_{157}$ from \citet{2018AJ....155...56J}, 2011~HM$_{102}$ from \citet{2013AJ....145...96P}, 2008~LC$_{18}$ from \citet{Bolin2023}, and others from \citet{Markwardt2023}.}
\label{fig:optical+NIR}
\end{figure}

The wavelength range of the NIRSpec instrument overlaps with common optical photometric bands. To convert such observations, we subtract solar colors in each filter \citep{Holmberg2006} and calculate reflectance values using the following equation:
\begin{equation}
f_r = 10^{0.4[(g-r)-(g-r)_{\astrosun}]}. 
\end{equation}
Here $f_r$ is the relative reflectance in the r filter, $(g-r)$ is the color of the object, and $(g-r)_{\astrosun}$ is the color of the Sun. In this case, g filter is the reference (i.e., $f_g=1$), and $f_r$ is relative to the reflectance at g. The g-r color can be replaced with any other color. Figure~\ref{fig:optical+NIR} shows that after converting the optical colors to the reflectance, they are generally consistent with the shortest wavelengths of the JWST observations for all of our targets. Therefore, we calculate the optical slope, S' using the gri colors (central wavelength 0.477, 0.623, and 0.763 $\mu$m, respectively) or BVR colors (central wavelength 0.445, 0.551, and 0.658 $\mu$m, respectively) with these equations:

\begin{eqnarray*}
S'_{gr} = (f_r - fg)/(\lambda_r - \lambda_g)\\
S'_{gi} = (f_i - fg)/(\lambda_i - \lambda_g),
\end{eqnarray*}
where $\lambda$ is the central wavelength of each filter. Similar equations can also be derived for BVR color. We then take the average of these slopes to be S'. 
Moreover, we calculated the near-IR slope, SIR$_1$, using the 0.7 to 1.2 $\mu$m spectra following the same procedure as \citet{disco2024}. We note that a similar near-IR slope could also be obtained from the photometry, since Figure~\ref{fig:optical+NIR} shows the i- and z-band photometry follow the spectra fairly well.

\begin{figure}
\plotone{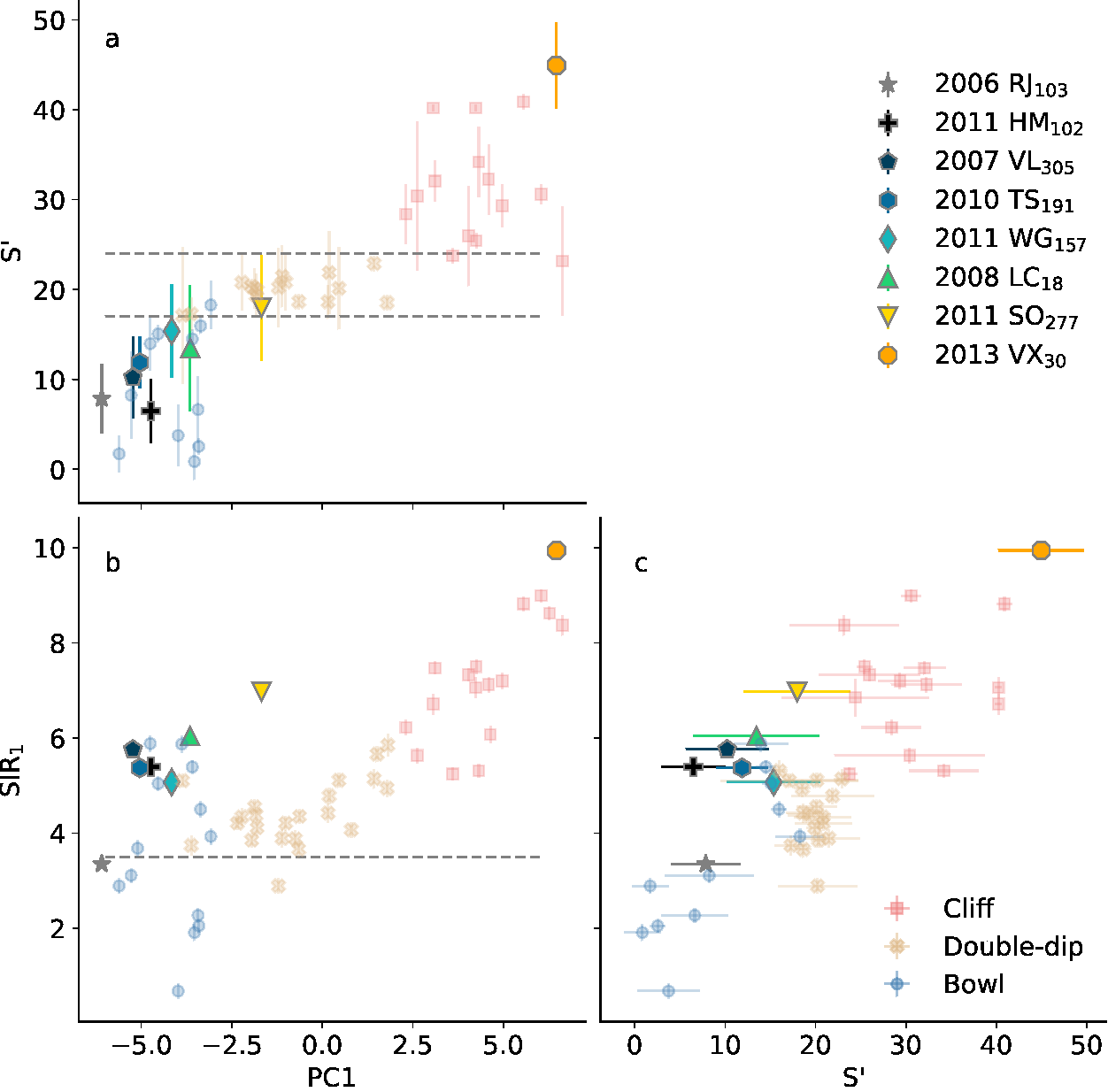}
\caption{PC1 versus optical (S') and near-IR slope (SIR$_1$). Each NT is depicted with a unique color marker in the plots. The dashed lines in the upper left panel (a) demark S' = 17 and 24 $\%/1000$\r{A}, our proposed borders to separate the three major spectral groups. The dashed line in the bottom left panel (b) demark SIR$_1$ = 3.5 $\%/1000$\r{A}, our proposed boarder to separate the shallow NIR and steep NIR slope subtypes of Bowl type object.}
\label{fig:slope_pc}
\end{figure}

In Section~\ref{sec:jwst}, we used a PCA to classify NT and TNO spectra into three major groups and identify outliers (see Figure~\ref{fig:PCA}). By the nature of PCA, the first principal component, PC1, captures most of the information from a spectrum into one single value. Therefore, we argue that PC1 is undoubtedly tied to the bulk surface composition of the object. In fact, Figure~\ref{fig:PCA} shows that PC1 alone can roughly distinguish the three major groups. To make this point clearer, Figure~\ref{fig:slope_pc} shows these PC1 values plotted against the optical (S') and near-IR slopes (SIR$_1$) of both the NTs and TNOs. The NT data for this plot is included in Table~\ref{Tab:slopes}, and the TNO data can be found in \citet{disco2024}. Figure~\ref{fig:slope_pc} clearly shows that PC1 and S' are correlated, and we can roughly distinguish the three major spectra groups with S' only, such as:
\begin{enumerate}
  \item[] S' $\lesssim 17~\%/1000$\r{A}: Bowl type,
  \item[] $17 \lesssim$ S' $\lesssim 24~\%/1000$\r{A}: Double-dip type,
  \item[] S' $\gtrsim 24~\%/1000$\r{A}: Cliff type.
\end{enumerate}
In this classification scheme, only 1 of the 16 Bowl types and 2 of the 15 Cliff types have been misidentified as Double-dip spectra. Notably, 2011 SO$_{277}$ would also be classified as a Double-dip type despite being a clear outlier in the PC1 vs. PC2 space (see Figure~\ref{fig:PCA}). However, our bluest and reddest objects, 2006~RJ$_{103}$ and 2013~VX$_{30}$, still lie at the far ends of the Bowl and Cliff groups respectively.

SIR$_1$ is also somewhat correlated with PC1 (see the lower left panel in Figure~\ref{fig:slope_pc}). While one can easily separate the Cliff types in this case, there is overlap between the Bowls and Double-dips. However, unlike S', the outlier 2011~SO$_{277}$ can be distinguished from the Double-dips due to its much larger SIR$_1$ value. Interestingly, the Bowl types have the largest span in SIR$_1$ of any of the spectral groups and could even be separated into shallow NIR slope and steep NIR slope subtypes. If we define the shallow NIR slope group as mostly distinguishable from the Double-dip objects via SIR$_1$, then the boundary is SIR$_1$ $\lesssim$ 3.5 $\%/1000$\r{A}. Except 2006~RJ$_{103}$, all other Bowl-type NTs belong to the steep NIR slope group.

Since S' generally traces PC1 and SIR$_1$ can distinguish the outlier 2011~SO$_{277}$, we plot them against each other in the lower right panel of Figure~\ref{fig:slope_pc}. Here, the three spectral groups, as well as the outlier, are separated in an optical/NIR space. Recall, S' and SIR$_1$ are both observables of many optical facilities, and this diagram is analogous to color-color diagrams, such as g-r vs r-J presented in \citet{Fraser2023} or g-r vs r-z presented in \citet{Bernardinelli2025}. We note that the S' vs. SIR$_{1}$ diagram in Figure~\ref{fig:slope_pc} is quite similar to the g-r vs. r-z (or r-i) plot in Figure 1 of \citet{Bernardinelli2025}.

From the above results, we present here a possible schema to map the optical/near-IR colors to the spectral groups.  As \citet{Fraser2023} and \citet{Bernardinelli2025} demonstrate, TNOs can be grouped as near-IR bright (Bright-IR in \citealt{Fraser2023}, NIRB in \citealt{Bernardinelli2025}) and near-IR faint (Faint-IR in \citealt{Fraser2023}, NIRF in \citealt{Bernardinelli2025}) in the optical/near-IR color space. Here we use the \citet{Bernardinelli2025} definition (but a similar schema should also be applicable to the \citealt{Fraser2023} definition). These two groups are more distinguishable in visible color (i.e., g-r) than in near-IR color (i.e., r-z; see the histogram in Figure 1 of \citealt{Bernardinelli2025}). We see the same phenomenon in Figure~\ref{fig:slope_pc} as S' can mostly distinguish the three spectral groups. Since NIRBs have the bluer optical colors of the two (see upper left panel of Figure 1 in \citet{Bernardinelli2025}), we predict that they would be primarily Bowl-type objects (though their spectral slope should be compared to our proposed borders in Figure~\ref{fig:slope_pc}). Notably, \citet{Bernardinelli2025} found that NTs are predominately NIRB, which is consistent with our results that NTs are primarily Bowl types and supports our proposed schema. On the other hand, the optically red NIRF are most likely red Cliff-types based on our classification scheme. \citet{Bernardinelli2025} found that all Cold Classical TNOs are NIRF and would be Cliff-types based on our scheme; this is consistent with the results in \citet{disco2024}. Other TNO populations were found to be $\sim$70\% NIRB \citep{Bernardinelli2025}, so we would expect a similar percentage of Bowl-type TNOs. However, Double-dip types were found to be the most common, at nearly 50\% of TNOs \citep{disco2024}. The values of g-r $\sim$ 0.75 and r-i $\sim$ 0.25 can mostly separate NIRB from NIRF in one-dimensional space \citep[see the histogram in Figure 1 of][]{Bernardinelli2025}, and these values are both roughly equal to S' $\sim 20~\%/1000$\r{A}, which is about the mean S' of Double-dip types. Therefore, the Double-dips should have g-r color very near 0.75, in between Bowls and Cliffs in optical color space.

One could then ask, are the Double-dips the ``reddest NIRBs'' or the ``bluest NIRFs''? As shown in the lower left panel of Figure~\ref{fig:slope_pc}, although the SIR$_{1}$ values of the Double-dip objects overlap with the Bowls, the Bowl types ultimately span a wider range of SIR$_{1}$. In particular, some Bowl-type objects have smaller SIR$_{1}$ values than \textit{any} Double-dip. Since the variation in near-IR slope is distinct to the Bowl-types, it seems reasonable to conclude that this population is somehow distinct from the Double-dips and Cliffs; therefore, we argue that Double-dips are more likely the bluest NIRFs. We should also note that \citet{Bernardinelli2025} also noticed increased variation in the NIRB group and argued that it could be further separated into NIRB+ and NIRB-, with Hot Classical TNOs being dominated by NIRB- and detached and scattering TNOS being dominated by NIRB+. Our shallow NIR slope and steep NIR slope subtypes are probably analogous.

Our outlier, 2011~SO$_{277}$, has the same optical color was Double-dips but with a much redder near-IR color (see lower right panel of Figure~\ref{fig:slope_pc}). Therefore, it could be considered the ``reddest NIRB'' or a complete outlier in this schema as well. Since we were not able to find objects with similar spectra in the TNO dataset (see Sec.~\ref{sec:comparsion}), it seems likely that 2011~SO$_{277}$ is a rare type of object. However, as more outer Solar System objects are discovered and have robust photometric observations taken through large-scale surveys like LSST, it might be worth specifically searching for 2011~SO$_{277}$-like objects near that color-color space to understand if there is another major spectral group not captured by current datasets.

While the color-to-spectra schema described above can generally distinguish the three main spectral groups, as well as outliers like 2011~SO$_{277}$, it does have limitations. This mapping is based on the correlation between optical/near-IR slopes and the principal spectral trends (PC1) of the three major groups. Consequently, objects with spectra significantly different from those in the \citet{disco2024} and NT samples, such as certain Centaurs in the \citet{discocentaur2024} sample, may not be accurately identified using this schema. Additionally, the boundaries between the three major spectral groups in the color-color space alone are not sharply defined, in particular between the Bowl and Double-dip types (see Fig. 5c), which could lead to misclassifications when using this method. Since Double-dips are distinguished by their particular strong CO$_2$ features, one solution to this particular problem could be additional photometric measurements around 3.7 $\mu$m and $\sim$ 4.2 $\mu$m, but this idea would need to be tested. Ultimately, spectral observations remain indispensable for determining the proper taxonomic classification of such objects as well as the identification and quantification of specific molecular ice species on their surfaces.

\section{Summary}\label{sec:summary}

In this work, we present the 0.7 - 5.0 $\mu$m spectra as measured by NIRSpec on JWST of eight Neptunian Trojans. Overall, we found clear variation in the surface composition of the NT population. Most NTs had features indicative of crystalline water ice, and all spectra had CO$_2$ ice but no clear CO ice. We found the spectrum of the reddest NT, 2013~VX$_{30}$, to be clearly distinct from the rest of the population with by far the strongest absorption between 3 and 4 $\mu$m (Figure~\ref{fig:images}). The bluest NT, 2006~RJ$_{103}$, appears to be depleted in water ice compared to most of the NT population. The other ultrared NT, 2011~SO$_{277}$ has similarly weak water features but a moderate amount of absorption due to aliphatic organics.

We also conducted a principal component analysis to compare our NTs to the \citet{disco2024} TNO sample and \citet{discocentaur2024} Centaur sample. Under their trimodal taxonomy (Bowl, Double-dip, and Cliff), we found that the majority of the NTs can be classified as Bowl-type spectra, while the reddest, 2013~VX$_{30}$, is a Cliff-type (Figure~\ref{fig:PCA}). While Double-dips were the most common type in the TNO sample, we find none amongst our NT sample. Instead, we identify two outliers: the bluest NT in our sample, 2006~RJ$_{103}$, falls very close to the Bowl-type group but has comparatively insignificant water absorption features, and the other red object, 2011~SO$_{277}$, falls completely outside this taxonomy system (see Figure \ref{fig:PCA}).

Except for the true outlier, 2011~SO$_{277}$, we are able to find the close counterparts for each of the NTs in the TNO spectra (Figure~\ref{fig:comp}). However, most of the counterparts are either Plutinos or ``distant" Centaurs (objects with perihelion $< 30$AU but semi-major axes $> 30$AU). Since these populations should experience different thermal environments than the NTs, these similarities point to a common primordial origin for scattered populations. Similarly, the counterpart for Cliff-type 2013~VX$_{30}$ is a Cliff-type Plutino not a Cold Classical TNO (which were found to be exclusively Cliff-types by \citet{disco2024}). The differences between 2013~VX$_{30}$ and Cold Classicals support the idea that Cold Classicals formed at their current location and therefore originate from a source population which is unique from scattered outer Solar System populations. The counterpart to the blue NT, 2006~RJ$_{103}$, is a similarly blue Plutino, 2004~EW$_{95}$. This Plutino has been suggested to be a former carbonaceous asteroid, while 2006~RJ$_{103}$ has been suggested to be comprised of hydrated silicates \citep{Sharkey2023}. Based on our results, we hypothesize that these kinds of objects could come from a region closer to the Sun than other primordial TNOs, with subsequent capture of CO$_2$ after being transported to the cooler outer Solar System. No clear counterpart for 2011~SO$_{277}$ could be found, further suggesting it is a truly unique object. Its similarities to water-depleted 2006~RJ$_{103}$ but also organic-rich 2013~VX$_{30}$ hint that it could be some sort of transitionary object. We propose that 2011~SO$_{277}$ could represent an irradiation and/or thermally-driven spectral state between neutral, water-rich Bowl types and neutral, spectrally featureless objects like 2006~RJ$_{103}$. However, it is still entirely unclear if and why objects like 2011~SO$_{277}$ are rare. Albedo observations could compliment these spectral observations and help disentangle the problem.

We also showed that the optical slope, S', is an effective tracer of the overall 0.7 - 5 $\mu$m spectral shape (i.e., PC1 value; see Figure~\ref{fig:optical+NIR}, upper left panel). The 0.7 - 1.2 $\mu$m near-IR slopes, SIR$_{1}$, which can also be measured via broad-band photometry, are also useful to distinguish the Cliff-type spectra from the rest and identify outliers similar to 2011~SO$_{277}$ (Figure~\ref{fig:optical+NIR}, lower left panel). Therefore, broad-band optical photometry appears to be a valid method to infer the surface compositions of NTs and outer Solar System small bodies. We advocate for using these S' cutoffs as preliminarily classifications of spectral type:  S' $\lesssim 17~\%/1000$\r{A} = Bowl type;  17 $\lesssim$\ S' $\lesssim 24~\%/1000$\r{A} = Double-dip type; S' $\gtrsim 24~\%/1000$\r{A} = Cliff type. We find that the Bowl-types could be further subdivided (and outliers identified) using SIR$_1$, similar to NIRB+ and NIRB- \citep{Bernardinelli2025}. We stress that this schema cannot detect specific spectral features/compositions. However, it has the potential to be particularly valuable for studying objects too faint for spectroscopic observations or for pritorizing targets for follow-up observations. We expect these findings to be highly relevant to large photometric surveys, such as the Dark Energy Survey \citep{Bernardinelli2025} and the Vera C. Rubin Observatory's LSST.

Notably, we did not carry out any theoretical fits to our observed spectra here to determine the precise surface composition of our targets. Doing so would require modeling the reflectance/scattering properties of the regolith on these objects (traditionally done using a Hapke model), which is beyond the scope of this work. Moreover, such models and current optical constants in the literature have not been able to sufficiently model similar spectra across this entire near-IR wavelength range \citep{disco2024}; further theoretical and laboratory work is needed. However, such compositional measurements would not only lead to a better understanding of the applicability of our conclusions but also of the formation and evolution of NTs, other outer Solar System small bodies, and the giant planets as a whole. Additionally, next-generation observatories, such as the Vera C. Rubin Observatory, are expected to dramatically increase our census and color inventory of small bodies in the Solar System \citep{2019EPSC...13.1224J}. These observations will lead to the best constrained color distributions of these populations to date and will determine how rare objects, such  as 2011~SO$_{277}$ and 2006~RJ$_{103}$, truly are in the outer Solar System. Future spectral observations of such newly discovered and unique objects will be key to understanding the true compositional diversity of these regions. Finally, studying the surfaces of these distant objects is much easier through nearby observations rather than remote sensing. While there are significant differences between the Jovian and Neptunian Trojan populations, the results from the flybys of the Lucy Mission will provide invaluable insights into what the surfaces of small, distant, resonant bodies actually look like \citep{2024SSRv..220...47O}.

\begin{acknowledgments}
We thank Noem\'{\i} Pinilla-Alonso and Rosario Brunetto for kindly providing us the TNO spectra.
We thank Mike Brown, Charles Proffitt, Ana Carolina De Souza Feliciano, and Matthew Belyakov for helpful conversations that greatly improved this paper. 
All of the data presented in this paper were obtained from the Mikulski Archive for Space Telescopes (MAST) at the Space Telescope Science Institute. The specific observations analyzed can be accessed via \dataset[https://doi.org/10.17909/j66s-pv96
]{https://doi.org/10.17909/j66s-pv96
}. STScI is operated by the Association of Universities for Research in Astronomy, Inc., under NASA contract NAS5–26555. Support to MAST for these data is provided by the NASA Office of Space Science via grant NAG5–7584 and by other grants and contracts.
Support for program \#2550 was provided by NASA through a grant from the Space Telescope Science Institute, which is operated by the Association of Universities for Research in Astronomy, Inc., under NASA contract NAS 5-03127. Markwardt was also supported by the Marsden Fund Council from Government funding, managed by Royal Society Te Apārangi.
\end{acknowledgments}

%

\vspace{5mm}
\facilities{JWST(NIRSpec)}


\software{astropy \citep{2013A&A...558A..33A,2018AJ....156..123A}, scipy \citep{2020SciPy-NMeth}, scikit-learn \citep{scikit-learn}, jwst \citep{2023zndo...8247246B}}

\bibliography{sample631}{}

\begin{thebibliography}{}
\expandafter\ifx\csname natexlab\endcsname\relax\def\natexlab#1{#1}\fi
\providecommand{\url}[1]{\href{#1}{#1}}
\providecommand{\dodoi}[1]{doi:~\href{http://doi.org/#1}{\nolinkurl{#1}}}
\providecommand{\doeprint}[1]{\href{http://ascl.net/#1}{\nolinkurl{http://ascl.net/#1}}}
\providecommand{\doarXiv}[1]{\href{https://arxiv.org/abs/#1}{\nolinkurl{https://arxiv.org/abs/#1}}}

\bibitem[{{Astropy Collaboration} {et~al.}(2013){Astropy Collaboration}, {Robitaille}, {Tollerud}, {Greenfield}, {Droettboom}, {Bray}, {Aldcroft}, {Davis}, {Ginsburg}, {Price-Whelan}, {Kerzendorf}, {Conley}, {Crighton}, {Barbary}, {Muna}, {Ferguson}, {Grollier}, {Parikh}, {Nair}, {Unther}, {Deil}, {Woillez}, {Conseil}, {Kramer}, {Turner}, {Singer}, {Fox}, {Weaver}, {Zabalza}, {Edwards}, {Azalee Bostroem}, {Burke}, {Casey}, {Crawford}, {Dencheva}, {Ely}, {Jenness}, {Labrie}, {Lim}, {Pierfederici}, {Pontzen}, {Ptak}, {Refsdal}, {Servillat}, \& {Streicher}}]{2013A&A...558A..33A}
{Astropy Collaboration}, {Robitaille}, T.~P., {Tollerud}, E.~J., {et~al.} 2013, \aap, 558, A33, \dodoi{10.1051/0004-6361/201322068}

\bibitem[{{Astropy Collaboration} {et~al.}(2018){Astropy Collaboration}, {Price-Whelan}, {Sip{\H{o}}cz}, {G{\"u}nther}, {Lim}, {Crawford}, {Conseil}, {Shupe}, {Craig}, {Dencheva}, {Ginsburg}, {VanderPlas}, {Bradley}, {P{\'e}rez-Su{\'a}rez}, {de Val-Borro}, {Aldcroft}, {Cruz}, {Robitaille}, {Tollerud}, {Ardelean}, {Babej}, {Bach}, {Bachetti}, {Bakanov}, {Bamford}, {Barentsen}, {Barmby}, {Baumbach}, {Berry}, {Biscani}, {Boquien}, {Bostroem}, {Bouma}, {Brammer}, {Bray}, {Breytenbach}, {Buddelmeijer}, {Burke}, {Calderone}, {Cano Rodr{\'\i}guez}, {Cara}, {Cardoso}, {Cheedella}, {Copin}, {Corrales}, {Crichton}, {D'Avella}, {Deil}, {Depagne}, {Dietrich}, {Donath}, {Droettboom}, {Earl}, {Erben}, {Fabbro}, {Ferreira}, {Finethy}, {Fox}, {Garrison}, {Gibbons}, {Goldstein}, {Gommers}, {Greco}, {Greenfield}, {Groener}, {Grollier}, {Hagen}, {Hirst}, {Homeier}, {Horton}, {Hosseinzadeh}, {Hu}, {Hunkeler}, {Ivezi{\'c}}, {Jain}, {Jenness}, {Kanarek}, {Kendrew}, {Kern}, {Kerzendorf}, {Khvalko}, {King}, {Kirkby}, {Kulkarni},
  {Kumar}, {Lee}, {Lenz}, {Littlefair}, {Ma}, {Macleod}, {Mastropietro}, {McCully}, {Montagnac}, {Morris}, {Mueller}, {Mumford}, {Muna}, {Murphy}, {Nelson}, {Nguyen}, {Ninan}, {N{\"o}the}, {Ogaz}, {Oh}, {Parejko}, {Parley}, {Pascual}, {Patil}, {Patil}, {Plunkett}, {Prochaska}, {Rastogi}, {Reddy Janga}, {Sabater}, {Sakurikar}, {Seifert}, {Sherbert}, {Sherwood-Taylor}, {Shih}, {Sick}, {Silbiger}, {Singanamalla}, {Singer}, {Sladen}, {Sooley}, {Sornarajah}, {Streicher}, {Teuben}, {Thomas}, {Tremblay}, {Turner}, {Terr{\'o}n}, {van Kerkwijk}, {de la Vega}, {Watkins}, {Weaver}, {Whitmore}, {Woillez}, {Zabalza}, \& {Astropy Contributors}}]{2018AJ....156..123A}
{Astropy Collaboration}, {Price-Whelan}, A.~M., {Sip{\H{o}}cz}, B.~M., {et~al.} 2018, \aj, 156, 123, \dodoi{10.3847/1538-3881/aabc4f}

\bibitem[{{Baratta} {et~al.}(1991){Baratta}, {Leto}, {Spinella}, {Strazzulla}, \& {Foti}}]{Baratta1991}
{Baratta}, G.~A., {Leto}, G., {Spinella}, F., {Strazzulla}, G., \& {Foti}, G. 1991, \aap, 252, 421

\bibitem[{{Bennett} {et~al.}(2013){Bennett}, {Pirim}, \& {Orlando}}]{Bennet2013}
{Bennett}, C.~J., {Pirim}, C., \& {Orlando}, T.~M. 2013, Chemical Reviews, 113, 9086, \dodoi{10.1021/cr400153k}

\bibitem[{{Bernardinelli} {et~al.}(2025){Bernardinelli}, {Bernstein}, {Abbott}, {Aguena}, {Allam}, {Brooks}, {Carnero Rosell}, {Carretero}, {da Costa}, {Pereira}, {Davis}, {De Vicente}, {Desai}, {Diehl}, {Doel}, {Everett}, {Flaugher}, {Frieman}, {Garc{\'\i}a-Bellido}, {Gaztanaga}, {Gruendl}, {Gutierrez}, {Herner}, {Hinton}, {Hollowood}, {Honscheid}, {James}, {Kuehn}, {Lahav}, {Lee}, {Marshall}, {Mena-Fern{\'a}ndez}, {Miquel}, {Myles}, {Plazas Malag{\'o}n}, {Samuroff}, {Sanchez}, {Santiago}, {Sevilla-Noarbe}, {Smith}, {Suchyta}, {Tarle}, {Tucker}, {Vikram}, {Walker}, \& {Weaverdyck}}]{Bernardinelli2025}
{Bernardinelli}, P.~H., {Bernstein}, G.~M., {Abbott}, T.~M.~C., {et~al.} 2025, arXiv e-prints, arXiv:2501.01551, \dodoi{10.48550/arXiv.2501.01551}

\bibitem[{{Bolin} {et~al.}(2023){Bolin}, {Fremling}, {Morbidelli}, {Noll}, {van Roestel}, {Deibert}, {Delbo}, {Gimeno}, {Heo}, {Lisse}, {Seccull}, \& {Suh}}]{Bolin2023}
{Bolin}, B.~T., {Fremling}, C., {Morbidelli}, A., {et~al.} 2023, \mnras, 521, L29, \dodoi{10.1093/mnrasl/slad018}

\bibitem[{{Bottke} {et~al.}(2023){Bottke}, {Vokrouhlick{\'y}}, {Marschall}, {Nesvorn{\'y}}, {Morbidelli}, {Deienno}, {Marchi}, {Dones}, \& {Levison}}]{bottke2023}
{Bottke}, W.~F., {Vokrouhlick{\'y}}, D., {Marschall}, R., {et~al.} 2023, \psj, 4, 168, \dodoi{10.3847/PSJ/ace7cd}

\bibitem[{{Brown} {et~al.}(2011){Brown}, {Schaller}, \& {Fraser}}]{Brown2011}
{Brown}, M.~E., {Schaller}, E.~L., \& {Fraser}, W.~C. 2011, \apjl, 739, L60, \dodoi{10.1088/2041-8205/739/2/L60}

\bibitem[{{Brunetto} {et~al.}(2006){Brunetto}, {Barucci}, {Dotto}, \& {Strazzulla}}]{Brunetto2006}
{Brunetto}, R., {Barucci}, M.~A., {Dotto}, E., \& {Strazzulla}, G. 2006, \apj, 644, 646, \dodoi{10.1086/503359}

\bibitem[{{Brunetto} {et~al.}(2025){Brunetto}, {H{\'e}nault}, {Cryan}, {Pinilla-Alonso}, {Emery}, {Guilbert-Lepoutre}, {Holler}, {McClure}, {M{\"u}ller}, {Pendleton}, {de Souza-Feliciano}, {Stansberry}, {Grundy}, {Peixinho}, {Strazzulla}, {Bannister}, {Cruikshank}, {Harvison}, {Licandro}, {Lorenzi}, {de Pr{\'a}}, \& {Schambeau}}]{Brunetto2025}
{Brunetto}, R., {H{\'e}nault}, E., {Cryan}, S., {et~al.} 2025, \apjl, 982, L8, \dodoi{10.3847/2041-8213/adb977}

\bibitem[{{Bushouse} {et~al.}(2023){Bushouse}, {Eisenhamer}, {Dencheva}, {Davies}, {Greenfield}, {Morrison}, {Hodge}, {Simon}, {Grumm}, {Droettboom}, {Slavich}, {Sosey}, {Pauly}, {Miller}, {Jedrzejewski}, {Hack}, {Davis}, {Crawford}, {Law}, {Gordon}, {Regan}, {Cara}, {MacDonald}, {Bradley}, {Shanahan}, {Jamieson}, {Teodoro}, \& {Williams}}]{2023zndo...8247246B}
{Bushouse}, H., {Eisenhamer}, J., {Dencheva}, N., {et~al.} 2023, {JWST Calibration Pipeline}, 1.11.4,  Zenodo, \dodoi{10.5281/zenodo.8247246}

\bibitem[{{Cruikshank} {et~al.}(1998){Cruikshank}, {Roush}, {Bartholomew}, {Geballe}, {Pendleton}, {White}, {Bell}, {Davies}, {Owen}, {de Bergh}, {Tholen}, {Bernstein}, {Brown}, {Tryka}, \& {Dalle Ore}}]{1998Icar..135..389C}
{Cruikshank}, D.~P., {Roush}, T.~L., {Bartholomew}, M.~J., {et~al.} 1998, \icarus, 135, 389, \dodoi{10.1006/icar.1998.5997}

\bibitem[{{{\'C}uk} {et~al.}(2012){{\'C}uk}, {Hamilton}, \& {Holman}}]{Cuk2012}
{{\'C}uk}, M., {Hamilton}, D.~P., \& {Holman}, M.~J. 2012, \mnras, 426, 3051, \dodoi{10.1111/j.1365-2966.2012.21964.x}

\bibitem[{{Dalle Ore} {et~al.}(2015){Dalle Ore}, {Barucci}, {Emery}, {Cruikshank}, {de Bergh}, {Roush}, {Perna}, {Merlin}, \& {Dalle Ore}}]{DalleOre2015}
{Dalle Ore}, C.~M., {Barucci}, M.~A., {Emery}, J.~P., {et~al.} 2015, \icarus, 252, 311, \dodoi{10.1016/j.icarus.2015.01.014}

\bibitem[{{de la Fuente Marcos} \& {de la Fuente Marcos}(2012)}]{2007rw10}
{de la Fuente Marcos}, C., \& {de la Fuente Marcos}, R. 2012, \aap, 545, L9, \dodoi{10.1051/0004-6361/201219931}

\bibitem[{{De Pr{\'a}} {et~al.}(2025){De Pr{\'a}}, {H{\'e}nault}, {Pinilla-Alonso}, {Holler}, {Brunetto}, {Stansberry}, {de Souza Feliciano}, {Carvano}, {Harvison}, {Licandro}, {M{\"u}ller}, {Peixinho}, {Lorenzi}, {Guilbert-Lepoutre}, {Bannister}, {Pendleton}, {Cruikshank}, {Schambeau}, {McClure}, \& {Emery}}]{DePra2024}
{De Pr{\'a}}, M.~N., {H{\'e}nault}, E., {Pinilla-Alonso}, N., {et~al.} 2025, Nature Astronomy, \dodoi{10.1038/s41550-024-02276-x}

\bibitem[{{Fam{\'a}} {et~al.}(2010){Fam{\'a}}, {Loeffler}, {Raut}, \& {Baragiola}}]{Fama2010}
{Fam{\'a}}, M., {Loeffler}, M.~J., {Raut}, U., \& {Baragiola}, R.~A. 2010, \icarus, 207, 314, \dodoi{10.1016/j.icarus.2009.11.001}

\bibitem[{{Fraser} {et~al.}(2023){Fraser}, {Pike}, {Marsset}, {Schwamb}, {Bannister}, {Buchanan}, {Kavelaars}, {Benecchi}, {Tan}, {Peixinho}, {Gwyn}, {Alexandersen}, {Chen}, {Gladman}, \& {Volk}}]{Fraser2023}
{Fraser}, W.~C., {Pike}, R.~E., {Marsset}, M., {et~al.} 2023, \psj, 4, 80, \dodoi{10.3847/PSJ/acc844}

\bibitem[{{Gomes} {et~al.}(2005){Gomes}, {Levison}, {Tsiganis}, \& {Morbidelli}}]{Gomes2005}
{Gomes}, R., {Levison}, H.~F., {Tsiganis}, K., \& {Morbidelli}, A. 2005, \nat, 435, 466, \dodoi{10.1038/nature03676}

\bibitem[{{Gomes} \& {Nesvorn{\'y}}(2016)}]{Gomes2016}
{Gomes}, R., \& {Nesvorn{\'y}}, D. 2016, \aap, 592, A146, \dodoi{10.1051/0004-6361/201527757}

\bibitem[{{Gordon} {et~al.}(2022){Gordon}, {Bohlin}, {Sloan}, {Rieke}, {Volk}, {Boyer}, {Muzerolle}, {Schlawin}, {Deustua}, {Hines}, {Kraemer}, {Mullally}, \& {Su}}]{2022AJ....163..267G}
{Gordon}, K.~D., {Bohlin}, R., {Sloan}, G.~C., {et~al.} 2022, \aj, 163, 267, \dodoi{10.3847/1538-3881/ac66dc}

\bibitem[{{Grundy} {et~al.}(2024){Grundy}, {Tegler}, {Steckloff}, {Tan}, {Loeffler}, {Jasko}, {Koga}, {Blakley}, {Raposa}, {Engle}, {Thieberger}, {Hanley}, {Lindberg}, {Gomez}, \& {Madden-Watson}}]{Grundy2024}
{Grundy}, W.~M., {Tegler}, S.~C., {Steckloff}, J.~K., {et~al.} 2024, \icarus, 410, 115767, \dodoi{10.1016/j.icarus.2023.115767}

\bibitem[{{Hainaut} {et~al.}(2012){Hainaut}, {Boehnhardt}, \& {Protopapa}}]{Hainaut2012}
{Hainaut}, O.~R., {Boehnhardt}, H., \& {Protopapa}, S. 2012, \aap, 546, A115, \dodoi{10.1051/0004-6361/201219566}

\bibitem[{{Hainaut} \& {Delsanti}(2002)}]{Hainaut2002}
{Hainaut}, O.~R., \& {Delsanti}, A.~C. 2002, \aap, 389, 641, \dodoi{10.1051/0004-6361:20020431}

\bibitem[{{H{\'e}nault} {et~al.}(2025){H{\'e}nault}, {Brunetto}, {Pinilla-Alonso}, {Baklouti}, {Djouadi}, {Guilbert-Lepoutre}, {M{\"u}ller}, {Cryan}, {de Souza-Feliciano}, {Holler}, {de Pr{\'a}}, {Emery}, {McClure}, {Schambeau}, {Pendleton}, {Harvison}, {Licandro}, {Lorenzi}, {Cruikshank}, {Peixinho}, {Bannister}, \& {Stansberry}}]{Henault2025}
{H{\'e}nault}, E., {Brunetto}, R., {Pinilla-Alonso}, N., {et~al.} 2025, \aap, 694, A126, \dodoi{10.1051/0004-6361/202452321}

\bibitem[{{Holmberg} {et~al.}(2006){Holmberg}, {Flynn}, \& {Portinari}}]{Holmberg2006}
{Holmberg}, J., {Flynn}, C., \& {Portinari}, L. 2006, \mnras, 367, 449, \dodoi{10.1111/j.1365-2966.2005.09832.x}

\bibitem[{{Holt} {et~al.}(2020){Holt}, {Nesvorn{\'y}}, {Horner}, {King}, {Marschall}, {Kamrowski}, {Carter}, {Brookshaw}, \& {Tylor}}]{Holt2020}
{Holt}, T.~R., {Nesvorn{\'y}}, D., {Horner}, J., {et~al.} 2020, \mnras, 495, 4085, \dodoi{10.1093/mnras/staa1348}

\bibitem[{{Jewitt}(2009)}]{Jewitt2009AJ}
{Jewitt}, D. 2009, \aj, 137, 4296, \dodoi{10.1088/0004-6256/137/5/4296}

\bibitem[{{Jewitt}(2015)}]{2015AJ....150..201J}
---. 2015, \aj, 150, 201, \dodoi{10.1088/0004-6256/150/6/201}

\bibitem[{{Jewitt}(2018)}]{2018AJ....155...56J}
---. 2018, \aj, 155, 56, \dodoi{10.3847/1538-3881/aaa1a4}

\bibitem[{{Juric} {et~al.}(2019){Juric}, {Jones}, {Eggl}, {Moeyens}, {Ivezic}, \& {Schwamb}}]{2019EPSC...13.1224J}
{Juric}, M., {Jones}, R.~L., {Eggl}, S., {et~al.} 2019, in EPSC-DPS Joint Meeting 2019, Vol. 2019, EPSC--DPS2019--1224

\bibitem[{{Ka{\v{n}}uchov{\'a}} {et~al.}(2012){Ka{\v{n}}uchov{\'a}}, {Brunetto}, {Melita}, \& {Strazzulla}}]{Kavuchova2012}
{Ka{\v{n}}uchov{\'a}}, Z., {Brunetto}, R., {Melita}, M., \& {Strazzulla}, G. 2012, \icarus, 221, 12, \dodoi{10.1016/j.icarus.2012.06.043}

\bibitem[{{Kortenkamp} {et~al.}(2004){Kortenkamp}, {Malhotra}, \& {Michtchenko}}]{Kortenkamp2004}
{Kortenkamp}, S.~J., {Malhotra}, R., \& {Michtchenko}, T. 2004, \icarus, 167, 347, \dodoi{10.1016/j.icarus.2003.09.021}

\bibitem[{{Kouchi} \& {Kuroda}(1990)}]{Kouchi1990}
{Kouchi}, A., \& {Kuroda}, T. 1990, \nat, 344, 134, \dodoi{10.1038/344134a0}

\bibitem[{{Leto} \& {Baratta}(2003)}]{Leto2003}
{Leto}, G., \& {Baratta}, G.~A. 2003, \aap, 397, 7, \dodoi{10.1051/0004-6361:20021473}

\bibitem[{{Licandro} {et~al.}(2025){Licandro}, {Pinilla-Alonso}, {Holler}, {De Pr{\'a}}, {Melita}, {de Souza Feliciano}, {Brunetto}, {Guilbert-Lepoutre}, {H{\'e}nault}, {Lorenzi}, {Stansberry}, {Schambeau}, {Harvison}, {Pendleton}, {Cruikshank}, {M{\"u}ller}, {McClure}, {Emery}, {Peixinho}, {Bannister}, \& {Wong}}]{discocentaur2024}
{Licandro}, J., {Pinilla-Alonso}, N., {Holler}, B.~J., {et~al.} 2025, Nature Astronomy, 9, 245, \dodoi{10.1038/s41550-024-02417-2}

\bibitem[{{Lin} {et~al.}(2022){Lin}, {Markwardt}, {Napier}, {Adams}, \& {Gerdes}}]{Lin2022}
{Lin}, H.-W., {Markwardt}, L., {Napier}, K.~J., {Adams}, F.~C., \& {Gerdes}, D.~W. 2022, Research Notes of the American Astronomical Society, 6, 79, \dodoi{10.3847/2515-5172/ac6752}

\bibitem[{{Lin} {et~al.}(2016){Lin}, {Chen}, {Holman}, {Ip}, {Payne}, {Lacerda}, {Fraser}, {Gerdes}, {Bieryla}, {Sie}, {Chen}, {Burgett}, {Denneau}, {Jedicke}, {Kaiser}, {Magnier}, {Tonry}, {Wainscoat}, \& {Waters}}]{Lin2016}
{Lin}, H.~W., {Chen}, Y.-T., {Holman}, M.~J., {et~al.} 2016, \aj, 152, 147, \dodoi{10.3847/0004-6256/152/5/147}

\bibitem[{{Lin} {et~al.}(2019){Lin}, {Gerdes}, {Hamilton}, {Adams}, {Bernstein}, {Sako}, {Bernadinelli}, {Tucker}, {Allam}, {Becker}, {Khain}, {Markwardt}, {Franson}, {Abbott}, {Annis}, {Avila}, {Brooks}, {Carnero Rosell}, {Carrasco Kind}, {Cunha}, {D'Andrea}, {da Costa}, {De Vicente}, {Doel}, {Eifler}, {Flaugher}, {Garc{\'\i}a-Bellido}, {Hollowood}, {Honscheid}, {James}, {Kuehn}, {Kuropatkin}, {Maia}, {Marshall}, {Miquel}, {Plazas}, {Romer}, {Sanchez}, {Scarpine}, {Sevilla-Noarbe}, {Smith}, {Smith}, {Soares-Santos}, {Sobreira}, {Suchyta}, {Tarle}, {Walker}, \& {Wester}}]{Lin2019}
{Lin}, H.~W., {Gerdes}, D.~W., {Hamilton}, S.~J., {et~al.} 2019, \icarus, 321, 426, \dodoi{10.1016/j.icarus.2018.12.006}

\bibitem[{{Lin} {et~al.}(2021){Lin}, {Chen}, {Volk}, {Gladman}, {Murray-Clay}, {Alexandersen}, {Bannister}, {Lawler}, {Ip}, {Lykawka}, {Kavelaars}, {Gwyn}, \& {Petit}}]{Lin2021}
{Lin}, H.~W., {Chen}, Y.-T., {Volk}, K., {et~al.} 2021, \icarus, 361, 114391, \dodoi{10.1016/j.icarus.2021.114391}

\bibitem[{{Lykawka} {et~al.}(2011){Lykawka}, {Horner}, {Jones}, \& {Mukai}}]{Lykawka2011}
{Lykawka}, P.~S., {Horner}, J., {Jones}, B.~W., \& {Mukai}, T. 2011, \mnras, 412, 537, \dodoi{10.1111/j.1365-2966.2010.17936.x}

\bibitem[{{Markwardt} {et~al.}(2023){Markwardt}, {Wen Lin}, {Gerdes}, \& {Adams}}]{Markwardt2023}
{Markwardt}, L., {Wen Lin}, H., {Gerdes}, D., \& {Adams}, F.~C. 2023, \psj, 4, 135, \dodoi{10.3847/PSJ/ace528}

\bibitem[{{Mastrapa} {et~al.}(2009){Mastrapa}, {Sandford}, {Roush}, {Cruikshank}, \& {Dalle Ore}}]{2009ApJ...701.1347M}
{Mastrapa}, R.~M., {Sandford}, S.~A., {Roush}, T.~L., {Cruikshank}, D.~P., \& {Dalle Ore}, C.~M. 2009, \apj, 701, 1347, \dodoi{10.1088/0004-637X/701/2/1347}

\bibitem[{{Mill{\'a}n} {et~al.}(2024){Mill{\'a}n}, {Luna}, {Domingo}, {Santonja}, \& {Satorre}}]{Millan2024}
{Mill{\'a}n}, C., {Luna}, R., {Domingo}, M., {Santonja}, C., \& {Satorre}, M.~{\'A}. 2024, \apj, 970, 117, \dodoi{10.3847/1538-4357/ad4c67}

\bibitem[{{Morbidelli} {et~al.}(2005){Morbidelli}, {Levison}, {Tsiganis}, \& {Gomes}}]{Morbidelli2005}
{Morbidelli}, A., {Levison}, H.~F., {Tsiganis}, K., \& {Gomes}, R. 2005, \nat, 435, 462, \dodoi{10.1038/nature03540}

\bibitem[{{Nesvorn{\'y}} \& {Vokrouhlick{\'y}}(2009)}]{Nesvorny2009}
{Nesvorn{\'y}}, D., \& {Vokrouhlick{\'y}}, D. 2009, \aj, 137, 5003, \dodoi{10.1088/0004-6256/137/6/5003}

\bibitem[{{Nesvorn{\'y}} {et~al.}(2013){Nesvorn{\'y}}, {Vokrouhlick{\'y}}, \& {Morbidelli}}]{Nesvorny2013}
{Nesvorn{\'y}}, D., {Vokrouhlick{\'y}}, D., \& {Morbidelli}, A. 2013, \apj, 768, 45, \dodoi{10.1088/0004-637X/768/1/45}

\bibitem[{{Noll} {et~al.}(2012){Noll}, {Grundy}, {Schlichting}, {Murray-Clay}, \& {Benecchi}}]{huya2012}
{Noll}, K.~S., {Grundy}, W.~M., {Schlichting}, H., {Murray-Clay}, R., \& {Benecchi}, S.~D. 2012, \iaucirc, 9253, 2

\bibitem[{{Olkin} {et~al.}(2024){Olkin}, {Vincent}, {Adam}, {Berry}, {Englander}, {Gray}, {Levison}, {Salmon}, {Spencer}, {Stanbridge}, \& {Sutter}}]{2024SSRv..220...47O}
{Olkin}, C., {Vincent}, M., {Adam}, C., {et~al.} 2024, \ssr, 220, 47, \dodoi{10.1007/s11214-024-01082-1}

\bibitem[{{Parker}(2015)}]{Parker2015}
{Parker}, A.~H. 2015, \icarus, 247, 112, \dodoi{10.1016/j.icarus.2014.09.043}

\bibitem[{{Parker} {et~al.}(2013){Parker}, {Buie}, {Osip}, {Gwyn}, {Holman}, {Borncamp}, {Spencer}, {Benecchi}, {Binzel}, {DeMeo}, {Fabbro}, {Fuentes}, {Gay}, {Kavelaars}, {McLeod}, {Petit}, {Sheppard}, {Stern}, {Tholen}, {Trilling}, {Ragozzine}, {Wasserman}, \& {Ice Hunters}}]{2013AJ....145...96P}
{Parker}, A.~H., {Buie}, M.~W., {Osip}, D.~J., {et~al.} 2013, \aj, 145, 96, \dodoi{10.1088/0004-6256/145/4/96}

\bibitem[{Pedregosa {et~al.}(2011)Pedregosa, Varoquaux, Gramfort, Michel, Thirion, Grisel, Blondel, Prettenhofer, Weiss, Dubourg, Vanderplas, Passos, Cournapeau, Brucher, Perrot, \& Duchesnay}]{scikit-learn}
Pedregosa, F., Varoquaux, G., Gramfort, A., {et~al.} 2011, Journal of Machine Learning Research, 12, 2825

\bibitem[{{Peixinho} {et~al.}(2003){Peixinho}, {Doressoundiram}, {Delsanti}, {Boehnhardt}, {Barucci}, \& {Belskaya}}]{Peixinho2003}
{Peixinho}, N., {Doressoundiram}, A., {Delsanti}, A., {et~al.} 2003, \aap, 410, L29, \dodoi{10.1051/0004-6361:20031420}

\bibitem[{{Pinilla-Alonso} {et~al.}(2024){Pinilla-Alonso}, {Licandro}, {Brunetto}, {Henault}, {Schambeau}, {Guilbert-Lepoutre}, {Stansberry}, {Wong}, {Lunine}, {Holler}, {Emery}, {Protopapa}, {Cook}, {Hammel}, {Villanueva}, {Milam}, {Cruikshank}, \& {de Souza-Feliciano}}]{2024A&A...692L..11P}
{Pinilla-Alonso}, N., {Licandro}, J., {Brunetto}, R., {et~al.} 2024, \aap, 692, L11, \dodoi{10.1051/0004-6361/202450124}

\bibitem[{{Pinilla-Alonso} {et~al.}(2025){Pinilla-Alonso}, {Brunetto}, {De Pr{\'a}}, {Holler}, {H{\'e}nault}, {Feliciano}, {Lorenzi}, {Pendleton}, {Cruikshank}, {M{\"u}ller}, {Stansberry}, {Emery}, {Schambeau}, {Licandro}, {Harvison}, {McClure}, {Guilbert-Lepoutre}, {Peixinho}, {Bannister}, \& {Wong}}]{disco2024}
{Pinilla-Alonso}, N., {Brunetto}, R., {De Pr{\'a}}, M.~N., {et~al.} 2025, Nature Astronomy, 9, 230, \dodoi{10.1038/s41550-024-02433-2}

\bibitem[{{Poston} {et~al.}(2018){Poston}, {Mahjoub}, {Ehlmann}, {Blacksberg}, {Brown}, {Carlson}, {Eiler}, {Hand}, {Hodyss}, \& {Wong}}]{Poston2018}
{Poston}, M.~J., {Mahjoub}, A., {Ehlmann}, B.~L., {et~al.} 2018, \apj, 856, 124, \dodoi{10.3847/1538-4357/aab1f1}

\bibitem[{{Press} {et~al.}(2007){Press}, {G{\"o}tzinger}, {Reitzenstein}, {Hofmann}, {L{\"o}ffler}, {Kamp}, {Forchel}, \& {Yamamoto}}]{2007PhRvL..98k7402P}
{Press}, D., {G{\"o}tzinger}, S., {Reitzenstein}, S., {et~al.} 2007, \prl, 98, 117402, \dodoi{10.1103/PhysRevLett.98.117402}

\bibitem[{Seabold \& Perktold(2010)}]{seabold2010statsmodels}
Seabold, S., \& Perktold, J. 2010, in 9th Python in Science Conference

\bibitem[{{Seccull} {et~al.}(2018){Seccull}, {Fraser}, {Puzia}, {Brown}, \& {Sch{\"o}nebeck}}]{Seccull2018}
{Seccull}, T., {Fraser}, W.~C., {Puzia}, T.~H., {Brown}, M.~E., \& {Sch{\"o}nebeck}, F. 2018, \apjl, 855, L26, \dodoi{10.3847/2041-8213/aab3dc}

\bibitem[{{Sharkey} {et~al.}(2023){Sharkey}, {Reddy}, {Kuhn}, {Sanchez}, \& {Bottke}}]{Sharkey2023}
{Sharkey}, B. N.~L., {Reddy}, V., {Kuhn}, O., {Sanchez}, J.~A., \& {Bottke}, W.~F. 2023, \psj, 4, 223, \dodoi{10.3847/PSJ/ad0845}

\bibitem[{{Sheppard} \& {Trujillo}(2006)}]{Sheppard2006}
{Sheppard}, S.~S., \& {Trujillo}, C.~A. 2006, Science, 313, 511, \dodoi{10.1126/science.1127173}

\bibitem[{{Souza-Feliciano} {et~al.}(2024){Souza-Feliciano}, {Holler}, {Pinilla-Alonso}, {De Pr{\'a}}, {Brunetto}, {M{\"u}ller}, {Stansberry}, {Licandro}, {Emery}, {Henault}, {Guilbert-Lepoutre}, {Pendleton}, {Cruikshank}, {Schambeau}, {Bannister}, {Peixinho}, {McClure}, {Harvison}, \& {Lorenzi}}]{Souza-Feliciano2024}
{Souza-Feliciano}, A.~C., {Holler}, B.~J., {Pinilla-Alonso}, N., {et~al.} 2024, \aap, 681, L17, \dodoi{10.1051/0004-6361/202348222}

\bibitem[{{Thompson} {et~al.}(1987){Thompson}, {Murray}, {Khare}, \& {Sagan}}]{1987JGR....9214933T}
{Thompson}, W.~R., {Murray}, B.~G.~J.~P.~T., {Khare}, B.~N., \& {Sagan}, C. 1987, \jgr, 92, 14933, \dodoi{10.1029/JA092iA13p14933}

\bibitem[{{Tsiganis} {et~al.}(2005){Tsiganis}, {Gomes}, {Morbidelli}, \& {Levison}}]{Nice}
{Tsiganis}, K., {Gomes}, R., {Morbidelli}, A., \& {Levison}, H.~F. 2005, \nat, 435, 459, \dodoi{10.1038/nature03539}

\bibitem[{Virtanen {et~al.}(2020)Virtanen, Gommers, Oliphant, Haberland, Reddy, Cournapeau, Burovski, Peterson, Weckesser, Bright, {van der Walt}, Brett, Wilson, Millman, Mayorov, Nelson, Jones, Kern, Larson, Carey, Polat, Feng, Moore, {VanderPlas}, Laxalde, Perktold, Cimrman, Henriksen, Quintero, Harris, Archibald, Ribeiro, Pedregosa, {van Mulbregt}, \& {SciPy 1.0 Contributors}}]{2020SciPy-NMeth}
Virtanen, P., Gommers, R., Oliphant, T.~E., {et~al.} 2020, Nature Methods, 17, 261, \dodoi{10.1038/s41592-019-0686-2}

\bibitem[{{Wong} \& {Brown}(2015)}]{Wong2015}
{Wong}, I., \& {Brown}, M.~E. 2015, \aj, 150, 174, \dodoi{10.1088/0004-6256/150/6/174}

\bibitem[{{Wong} \& {Brown}(2016)}]{Wong2016}
---. 2016, \aj, 152, 90, \dodoi{10.3847/0004-6256/152/4/90}

\bibitem[{{Wong} \& {Brown}(2017)}]{Wong2017}
---. 2017, \aj, 153, 145, \dodoi{10.3847/1538-3881/aa60c3}

\bibitem[{{Wong} {et~al.}(2019){Wong}, {Mishra}, \& {Brown}}]{2019AJ....157..225W}
{Wong}, I., {Mishra}, A., \& {Brown}, M.~E. 2019, \aj, 157, 225, \dodoi{10.3847/1538-3881/ab1b22}

\bibitem[{{Wong} {et~al.}(2024){Wong}, {Brown}, {Emery}, {Binzel}, {Grundy}, {Marchi}, {Martin}, {Noll}, \& {Sunshine}}]{Wong2024}
{Wong}, I., {Brown}, M.~E., {Emery}, J.~P., {et~al.} 2024, \psj, 5, 87, \dodoi{10.3847/PSJ/ad2fc3}

\bibitem[{{Zhang} {et~al.}(2023){Zhang}, {Zhu}, {Turner}, {Antonov}, {Garcia}, {Meinert}, {Young}, {Jewitt}, \& {Kaiser}}]{Zhang2023}
{Zhang}, C., {Zhu}, C., {Turner}, A.~M., {et~al.} 2023, Science Advances, 9, eadg6936, \dodoi{10.1126/sciadv.adg6936}

\end{thebibliography}
\bibliographystyle{aasjournal}



\end{document}